\documentclass{aa}
\usepackage{epsfig,color}
\usepackage{natbib}
\bibpunct{(}{)}{;}{a}{}{,}
\def\cf{{\rm cf. }}

\def\spose#1{\hbox to 0pt{#1\hss}}
\def\ltsimm{\mathrel{\spose{\lower 3pt\hbox{$\sim$}}
        \raise 2.0pt\hbox{$<$}}}
\def\gtsimm{\mathrel{\spose{\lower 3pt\hbox{$\sim$}}
        \raise 2.0pt\hbox{$>$}}}
\def\Mdot{\hbox{${\dot M}$}}
\def\vinfty{\hbox{${v_{\infty}}$} \,}
\def\km{{\rm\thinspace km}}
\def\cm{{\rm\thinspace cm}}
\def\kpc{{\rm\thinspace kpc}}
\def\s{{\rm\thinspace s}}
\def\yr{{\rm\thinspace yr}}
\def\g{{\rm\thinspace g}}
\def\kmps{\hbox{${\rm\km\s^{-1}\,}$}}

\def\erg{{\rm\thinspace erg}}
\def\Hz{{\rm\thinspace Hz}}
\def\keV{{\rm\thinspace keV}}
\def\mJy{{\rm\thinspace mJy}}
\def\ster{{\rm\thinspace ster}}
\def\ergps{\hbox{${\rm\erg\s^{-1}\,}$}}
\def\Rsol{\hbox{${\rm\thinspace R_{\odot}}$}}
\def\Msol{\hbox{${\rm\thinspace M_{\odot}}$}}
\def\Lsol{\hbox{${\rm\thinspace L_{\odot}}$}}
\def\Msolpyr{\hbox{${\rm\Msol\yr^{-1}\,}$}}

\def\pcmc{\hbox{${\rm\cm^{-3}\,}$}}
\def\ergpscm3Hz{\hbox{${\rm\ergps\cm^{-3}\Hz^{-1}\,}$}}
\def\ergpscm3Hzster{\hbox{${\rm\ergps\cm^{-3}\Hz^{-1}\ster^{-1}\,}$}}
\def\gpcm3{\hbox{${\rm\g\cm^{-3}\,}$}}
\def\ergpcm2{\hbox{${\rm\erg\cm^{-2}\,}$}}
\def\ergpcm3{\hbox{${\rm\erg\cm^{-3}\,}$}}
\def\ergpscm2{\hbox{${\rm\ergps\cm^{-2}\,}$}}
\def\phpscm2{\hbox{${\rm photons\;\s\cm^{-2}\,}$}}
\def\wr{{\rm WR\thinspace}}

\def\aap{{\rm A\&A}}
\def\apj{{\rm ApJ}}
\def\apjs{{\rm ApJS}}
\def\aj{{\rm AJ}}
\def\mnras{{\rm MNRAS}}
\def\nat{{\rm Nature}}
\def\apss{{\rm Ap\&SS}}
\def\araa{{\rm ARA\&A}}
\def\physrep{{\rm PhR}}

\def\sva{{\rm Sov. Astr.}}
\def\nuclfu{{\rm Nucl. Fusion}}
\value{1}
\begin{document}

\title{Radio emission models of Colliding-Wind Binary Systems - Inclusion
of IC cooling}

\author{J.M.~Pittard\inst{1}, S.M.~Dougherty\inst{2}, R.F.~Coker\inst{3},
E.~O'Connor\inst{2,4} and N.J.~Bolingbroke\inst{2,5}}

\institute{School of Physics and Astronomy, The University of Leeds,
        Woodhouse Lane, Leeds LS2 9JT, UK \and National Research
        Council of Canada, Herzberg Institute for Astrophysics,
        Dominion Radio Astrophysical Observatory, P.O. Box 248,
        Penticton, BC, V2A 6J9, Canada \and Los Alamos
        National Laboratory, X-2 MS T-087, Los Alamos, NM 87545, USA
        \and Physics Department, University of Prince Edward Island,
        Charlottetown, PEI, Canada \and Department of Physics and
        Astronomy, University of Victoria, 3800 Finnerty Rd, Victoria,
        BC, Canada}

\offprints{J.M.~Pittard,\\\email{jmp@ast.leeds.ac.uk}}

\date{Accepted 3rd October, 2005}

\abstract{Radio emission models of colliding wind binaries (CWBs) have
been discussed by \cite{Dougherty:2003}.  We extend these models by
considering the temporal and spatial evolution of the energy
distribution of relativistic electrons as they advect downstream from
their shock acceleration site.  The energy spectrum evolves
significantly due to the strength of inverse-Compton (IC) cooling in
these systems, and a full numerical evaluation of the synchrotron
emission and absorption coefficients is made.  We have demonstrated
that the geometry of the WCR and the streamlines of the flow within it
lead to a spatially dependent break frequency in the synchrotron
emission. We therefore do not observe a single, sharp break in the
synchrotron spectrum integrated over the WCR, but rather a steepening
of the synchrotron spectrum towards higher frequencies. We also
observe that emission from the wind-collision region (WCR) may appear
brightest near the shocks, since the impact of IC cooling on the
non-thermal electron distribution is greatest near the contact
discontinuity (CD), and demonstrate that the impact of IC cooling on
the observed radio emission increases significantly with decreasing
binary separation. We study how the synchrotron emission changes in
response to departures from equipartition, and investigate how the
thermal flux from the WCR varies with binary separation. 
Since the emission from the WCR is optically thin, we 
see a substantial fraction of this emission at certain viewing angles,
and we show that the {\em thermal} emission from a CWB can mimic a thermal plus
non-thermal composite spectrum if the thermal emission from the WCR
becomes comparable to that from the unshocked winds.
We demonstrate that the observed synchrotron emission depends upon the
viewing angle and the wind-momentum ratio, and find that the observed
synchrotron emission decreases as the viewing angle moves through the
WCR from the WR shock to the O shock. We obtain a number of insights 
relevant to models of closer systems such as \object{\wr140}. Finally, we
apply our new models to the very wide system \object{\wr147}. The 
acceleration of non-thermal electrons appears to be very 
efficient in our models of \object{\wr147}, and we suggest that the shock structure
may be modified by feedback from the accelerated particles.
}

\titlerunning{Radio emission models of Colliding-Wind Binary Systems}
\authorrunning{J.M.~Pittard et al.}

\maketitle

\label{firstpage}

\section{Introduction}
\label{sec:intro}
Radio observations of early-type stars reveal that they can be sources
of both thermal and non-thermal radiation. The thermal emission is
readily explained as free-free emission within the stellar wind
\citep{Wright:1975}, while the non-thermal emission is attributed to
relativistic electrons moving in a magnetic field. The consensus is
that the relativistic electrons in the winds of massive stars are
created through diffusive shock acceleration (DSA), though other mechanisms
have been investigated \citep[e.g.,][]{Jardine:1996}. Observations of the WR+OB
binary systems \object{\wr140} \citep{Dougherty:2005}, \object{\wr146}
\citep{Dougherty:1996, Dougherty:2000a} and \object{\wr147} \citep{Moran:1989,
Churchwell:1992, Williams:1997, Niemela:1998} clearly show the
non-thermal emission in these binaries arises from a wind-collision
region (WCR) bounded by strong shocks formed where the two stellar
winds collide. There is strong evidence that all WR stars that
exhibit non-thermal emission are binary \citep{Dougherty:2000b}.

For synchrotron emission to be observed, the relativistic electrons
must be within the optically thin region of the stellar wind(s), which
in the case of massive stars implies at more than a few hundred
stellar radii from the underlying star.  These large radii are
problematic for single star models of synchrotron emission, where the
declining velocity jump and compression ratio with increasing radius
of wind-embedded shocks produces a rapidly decreasing synchrotron
emissivity \citep{vanLoo:2005b}.

CWB systems are important objects for investigating the underlying
physics of DSA because they provide access to
higher mass, radiation and magnetic field energy densities than is
possible through studies of supernova remnants.  However, until the
work of \cite{Dougherty:2003} (hereafter Paper~I), CWB models of radio
spectra were typically based on a highly simplified model consisting
of a thermal source plus a point-like source of non-thermal emission
attenuated by free-free absorption \citep[e.g.,][]{Chapman:1999,
Monnier:2002}. Use of these models often resulted in difficulties in
fitting observational data \citep[see, e.g.,][]{Williams:1990,
White:1995}.  As a first step toward the construction of more
realistic models with better predictive qualities, a hydrodynamical
model of the WCR was used in Paper~I to obtain a more accurate
representation of the spatial distribution of the free-free and
non-thermal emission from CWB systems. Several important scaling laws
were determined and, in spite of other simplifying assumptions, the
results of that work showed great promise in comparisons with
observed continuum spectra and images. We refer the reader to
Paper~I for essential background to the work presented here.

Inverse-Compton (hereafter IC) cooling is an important 
energy-loss mechanism for relativistic electrons in CWB
systems, since the energy distribution will evolve
as the electrons advect away from the shocks. The
inclusion of this process and the calculation of emission and
absorption coefficients from an arbitrary energy distribution of
non-thermal electrons are the main enhancements to the model and the
work presented in this paper. Following examination of the effect of
these mechanisms, we apply these new models to \object{\wr147}, and close by
describing future directions for this work.


\section{Modelling the radio emission from CWBs}
\label{sec:modelling}
In Paper~I, the radio emission and absorption from a CWB system was
calculated from a 2D axis-symmetric hydrodynamical model of the
stellar winds and the collision zone.  The
temperature and density values on the hydrodynamic grid were used to
calculate the free-free emission and absorption coefficients from each
grid cell. The synchrotron emission and self-absorption
from each cell within the WCR was calculated assuming that the
distribution of relativistic electrons in each cell could be specified
by a power-law, i.e. $n_{\rm e}(\gamma) \propto \gamma^{-p}$, where $\gamma$ is
the Lorentz factor\footnote{For a strong shock with a compression ratio of 4,
test particle theory predicts that $p=2$ (see references in Paper~I).}.
The synchrotron emissivity, $P_{\nu}$, also followed a
power-law ($P_{\nu} \propto \nu^{-(p-1)/2}$), which was assumed to be
valid over all frequency, $\nu$.
The flux and intensity distribution at a specified
frequency were then obtained from a radiative transfer calculation.

The hydrodynamical simulations of Paper~I do not provide direct 
information on the magnetic field or relativistic particle distribution, so 
the magnetic energy density $U_{\rm B}$, and the relativistic electron energy
density $U_{\rm rel}$, were set proportional to the
thermal particle internal energy density, $U_{\rm th}$:

\begin{equation}
\label{eq:b_en_dens}
U_{\rm B} = B^{2}/8 \pi = \zeta_{\rm B} U_{\rm th},
\end{equation}

\noindent and
\begin{equation}
\label{eq:nt_en_dens}
U_{\rm rel} = \int n_{\rm e}(\gamma) \gamma m_{\rm e}c^{2} 
d\gamma = \zeta_{\rm rel} U_{\rm th}.
\end{equation}

\noindent $U_{\rm th}={P\over{\Gamma-1}}$, with $P$ the gas pressure
and $\Gamma$ the adiabatic index, assumed to be $5/3$, as for an ideal
gas. $\zeta_{\rm B}$ and $\zeta_{\rm rel}$ are constants which
determine the amount of synchrotron emission and absorption, and the
Razin turnover frequency in our models. Since we did not attempt to
model the acceleration of the relativistic particles from basic
principles, the values of $\zeta_{\rm B}$ and $\zeta_{\rm rel}$ are
not determined in a self-consistent fashion. Instead, when modelling
specific systems, their values are chosen to best match the observed
radio emission.  Our previous models in Paper~I showed that if
$\zeta_{\rm B}=\zeta_{\rm rel}$, then values of around 1\% are
required \citep[see also][]{Mioduszewski:2001}. Unless otherwise
noted, $\zeta_{\rm B} = \zeta_{\rm rel} = \zeta$, though we relax this 
restriction in Sec.~\ref{sec:zeta_var} and in some models in 
Sec.~\ref{sec:wr147}.

\subsection{The hydrodynamical code}
In contrast to Paper~I, a hydrodynamical code with adaptive mesh 
refinement is used here~\citep{Falle:1996,Falle:1998}. Such
codes concentrate more cells where the flow is interesting, for
example near shocks, and fewer cells where it is uniform, resulting in
significantly shorter computational times. We use a small amount of
artificial viscosity in the calculations to prevent the development of
the carbuncle instability at the apex of the WCR. This numerical
artifact develops when shocks are locally aligned with the grid
\citep{Walder:1994,Leveque:1998}.

\subsection{The non-thermal electron energy spectrum}
\label{sec:nt_spectrum}
In this work, it is assumed that the acceleration of relativistic
electrons {\em at the shocks} bounding the WCR produces an energy
distribution with a power-law spectrum.  When particle acceleration is
relatively inefficient, this is only strictly true in the absence of
cooling. However, the resulting deviation from a pure power-law is
restricted to energies which are very near to the high energy cutoff
\citep{Chen:1991,vanLoo:2005}, and we ignore this effect in the
current work. In contrast, when DSA is efficient, non-linear
effects modify the pre-shock wind speeds over a length scale
comparable to the diffusion length of the energetic particles, and
slow them before they undergo a shock transition. Thus higher energy
particles with larger diffusion length scatter in regions where the
pre-shock wind speed is closer to the terminal speed of the stellar
wind, while low energy particles do not diffuse very far from the
subshock and scatter in regions where the wind speed is lower due to
its journey through much of the shock precursor. This speed profile
results in a concave upward curvature to the non-thermal particle
spectrum, where $p$ decreases with increasing energy \citep[][and
references therein]{Ellison:2004}. Efficient particle acceleration can
also flatten the non-thermal particle energy distribution i.e. $p <
2$.  Modelling this process is non-trivial and we make no attempt in
the current work to calculate these effects
self-consistently. However, we discuss it further and perform
calculations with a non-thermal particle spectrum where $p<2$ in
Sec.~\ref{sec:wr147}.

Relativistic electrons that scatter out of the shock with Lorentz
factor $\gamma \ltsimm 10^{5}$ are frozen into the post-shock flow
\citep{White:1985}. As they are advected out of the system, the
relativistic electrons lose energy through the action of several
processes, including ionic cooling, magneto-bremsstrahlung cooling,
and IC cooling.
Since IC cooling strongly dominates magneto-bremsstrahlung cooling in
this work (see Eq.~14 of Paper~I), we do not consider the latter here.
Cooling reduces the non-thermal electron energy density at a given
$\gamma$ and alters the energy spectrum of these electrons from a
simple power-law.  Thus, in this work we assume the non-thermal
electron energy spectrum is a power-law at the shocks, but let it
evolve as the electrons advect downstream from the shocks. 
As noted earlier, the inclusion of these
processes is the main enhancement to the model since Paper~I, and in
the following sub-sections we describe in detail our method of
implementation.

\subsubsection{Inverse Compton cooling}
\label{sec:ic_cooling}
The high UV luminosity of the stars in CWBs means that IC cooling of the
relativistic electrons in the WCR can be very efficient,
particularly for high $\gamma$ electrons. In the Thomson limit, 
the rate of energy loss of a
relativistic electron exposed to an isotropic distribution of photons
by IC scattering is~\citep{Rybicki:1979}

\begin{equation}
\label{eq:compt}
P_{\rm compt} = \frac{4}{3} \sigma_{T} c \gamma^{2} \beta^{2} U_{\rm ph},
\end{equation}

\noindent where $\sigma_{T}$ is the Thomson cross section,
$\beta=v/c=\sqrt{1 - 1/\gamma^{2}}$, $U_{\rm ph}$ is the energy density 
of photons, and where elastic scattering is assumed and quantum
effects on the cross section are neglected. 
For a single star and relativistic particles with $\beta \approx 1$, 
we can rewrite Eq.~\ref{eq:compt} as

\begin{equation}
\label{eq:ic_loss}
\left.\frac{d\gamma}{dt}\right|_{ic} = \frac{\sigma_{T} \gamma^{2}}
{3 \pi m_{\rm e} c^{2}} \frac{L}{r^{2}} =  
8.61 \times 10^{-20} \gamma^{2} f\;{\rm s^{-1}},
\end{equation}

\noindent where $L$ is the stellar luminosity, $r$ is the distance of
the particles from the star, and $f = {{L}\over{r^{2}}}\;(\ergpscm2)$
is a measure of the radiative flux on an electron at distance r.  For
a given electron, the reduction in $\gamma$ due to IC cooling can be
calculated if the integral of $ft$ along its streamline
is known. In a CWB, $f={{L_{\rm O}}\over{r_{\rm O}^{2}}} + {{L_{\rm
WR}}\over{r_{\rm WR}^{2}}}$ due to the radiation field of both stars.
The value of $ft$ in any given cell on the hydrodynamic grid can be
obtained by the specification of a suitable scalar in the code. It is
then a simple matter during post-processing to calculate the IC-cooled
relativistic energy spectrum at that location.

\begin{figure*}[t]
\vspace{4.6cm}
\includegraphics{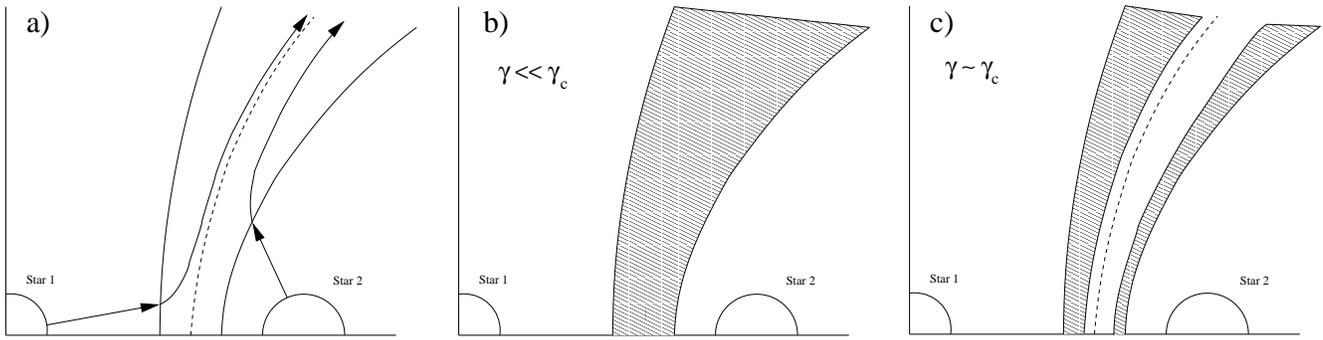}
\caption[]{Schematic diagrams of the flow within the WCR and
the distribution of relativistic electrons. a) The flow within 
the winds and the WCR, which is located nearer the star with the weaker
wind (star 2 in this example). The dotted line depicts the 
contact discontinuity (CD) between the two winds. Either side of the CD
is a shock which separates the high Mach number wind from high pressure 
post-shock gas. Example streamlines are shown for material leaving the stars
along specific direction vectors.
b) The distribution of relativistic electrons with 
Lorentz factor $\gamma << \gamma_{\rm c}$, assuming that such electrons 
are first-order Fermi accelerated at the shocks. c) As b) but for
$\gamma \sim \gamma_{\rm c}$. IC cooling effectively excludes electrons
with Lorentz factor $\gamma_{\rm c}$ from a region near the CD.}
\label{fig:iccool_schematic}
\end{figure*}

The influence of IC cooling increases in magnitude as plasma flows
from the shocks toward the contact discontinuity - this reflects the
increased exposure to the high intensity stellar radiation field
from the time when the relativistic electrons were accelerated
in one of the shocks bounding the WCR. In addition, since the rate of
cooling is proportional to $\gamma^{2}$, IC cooling is most
significant for electrons with the highest energies.  We can therefore
define a characteristic Lorentz factor, $\gamma_{\rm c}$, where IC
cooling becomes significant for flow near the apex of the WCR.  The flow
morphology and the distribution of electrons with a specific Lorentz
factor is shown in Fig.~\ref{fig:iccool_schematic}. For low-energy
electrons ($\gamma << \gamma_{\rm c}$), IC cooling is negligible and
such electrons may exist throughout the WCR, as shown in
Fig.~\ref{fig:iccool_schematic}b. In contrast, IC cooling is
significant for electrons with $\gamma \gtsimm \gamma_{\rm c}$, which
rapidly lose energy as they flow away from the shocks.  In such
circumstances, electrons with $\gamma \sim \gamma_{\rm c}$ are
effectively confined to narrow regions near the shocks, and absent
from the central volume of the WCR, as illustrated in
Fig.~\ref{fig:iccool_schematic}c.

\subsubsection{Ionic (coulomb) cooling}
\label{sec:ionic_cooling}
Relativistic electrons may also lose energy through Coulomb collisions
with thermal ions. The rate of energy loss through this process for
electrons of energy $\gamma$ is~\citep[\cf][]{Chen:1991}

\begin{equation}
\label{eq:ionic_cool}
\left.\frac{d\gamma}{dt}\right|_{\rm coul} = \frac{4 \pi e^{4} n_{\rm i}
\;{\rm ln\;} \Lambda}{m_{\rm e}^{2} c^{3} \sqrt{\gamma^{2} - 1}} = 
3.0\times 10^{-14}
\frac{n_{\rm i} \;{\rm ln\;}\Lambda} {\sqrt{\gamma^{2} - 1}}\;{\rm s}^{-1},
\end{equation}

\noindent where ${\rm ln}\;\Lambda = 29.7 + {\rm ln\;}n_{\rm i}^{-1/2}
T_{6}$ is the Coulomb logarithm
\citep[e.g.,][]{Cowie:1977}\footnote{While we use this expression for
the Coulomb logarithm, \cite{Li:2001} note that it should be
reduced to about half of its conventional value.}, $T_{6}$ is the
temperature in units of $10^{6}$~K, and $n_{\rm i}$ is the number density
of ions ($\pcmc$). As for IC cooling, we can estimate the degree of ionic
cooling, namely by constructing a scalar to track the value of $n_{\rm
i} t \;{\rm ln\;}\Lambda$ in each hydrodynamic cell of hot shocked
gas. Ionic coooling is greatest for non-thermal electrons with
relatively low energies, and while it is included in our model we note
that there is no observational signature of this process on the
synchrotron spectrum above 10~MHz (see
Fig~\ref{fig:standard_espec}). However, we include it for
completeness.

\subsubsection{Maximum energy of relativistic electrons}
\label{sec:gam_max}
In the standard DSA scheme, particles
are assumed to scatter from MHD waves in the background plasma, either
pre-existing or generated by the counter-streaming ions themselves
\citep{Bell:1978}.  
There is strong evidence that the efficiency of particle acceleration 
in the DSA scheme is dependent on the orientation of the upstream 
magnetic field to the shock normal \cite[see the review by][]{Blandford:1987}.
When the shock is quasi-perpendicular the competing ``shock drift'' 
mechanism may be more efficient \citep{Jokipii:1987}, though the
acceleration rates become equal in shocks with a high level of magnetic 
turbulence since these have no well-defined obliquity\footnote{The obliquity
of the shocks expected in CWB systems is described by \cite{Eichler:1993}.}.

Since the physics of electron injection is not yet fully understood,
we ignore these complications in the present work, and assume that
both the normalization of the non-thermal particle spectrum, and the
maximum energy that the accelerated particles can reach, are
independent of spatial position. To estimate the value of $\gamma_{\rm
max}$, we balance the rate of energy gain through 1st-order Fermi
acceleration, by the rate of energy loss due to IC cooling (the
dominant cooling mechanism for relativistic electrons with large
$\gamma$). The 1st-order Fermi power is
\citep{Chen:1991,vanLoo:2005}\footnote{\cite{Chen:1991} erroneously
assume that shock-accelerated electrons return to the shock after one
scattering, whereas electrons are actually scattered $c/2u$ times
before recrossing the shock~\citep{vanLoo:2005}.}
\begin{equation}
\left.\frac{d\gamma}{dt}\right|_{acc} = \frac{4}{3} \frac{\chi_{\rm
c}-1}{\chi_{\rm c}} \frac{u \gamma \beta^{2}}{\lambda} \frac{2u}{c},
\end{equation}
where $\chi_{\rm c}$ is the ratio of upstream to downstream flow speeds,
$u$ is the shock velocity, $c$ is the speed of light, and $\lambda$ is
the mean-free path of the accelerating charge. This latter quantity is
somewhat uncertain, but is typically estimated as several
gyro-radii. We adopt $\lambda \approx 3 r_{\rm g}$, where the
gyro-radius $r_{\rm g} = pc/qB$ \citep{White:1985}\footnote{This is
equivalent to specifying that the diffusion coefficient is
proportional to the particle energy.
}, and $p$ and $q$ are the momentum and charge of the accelerating particle.
Assuming that we are in the strong limit for adiabatic shocks
($\chi_{\rm c}=4$) and that the accelerated charge is a highly
relativistic electron ($\beta = 1$, $\gamma >> 1$), we have
\begin{equation}
\label{eq:fermi_gain}
\left.\frac{d\gamma}{dt}\right|_{acc} = \frac{1}{3} \frac{uqB}{m_{\rm e}c^{2}}
\frac{2u}{c}.
\end{equation} 
By balancing Eq.~\ref{eq:ic_loss} with Eq.~\ref{eq:fermi_gain} we obtain
\begin{equation}
\label{eq:gam_max}
\gamma_{\rm max} \approx 400 u r_{\rm O} \sqrt{\frac{B}{L_{\rm O}}}. 
\end{equation}
Eq.~\ref{eq:gam_max} is equivalent to Eq.~14 in \cite{Eichler:1993}.
Due to its dependence on $B$, $\gamma_{\rm max} \propto
\zeta_{\rm B}^{1/4}$. In our standard model, $\gamma_{\rm max} \approx
7 \times 10^{4}$ at the point where the shocks intersect the 
line of centers through the stars. However, since the value of
$\zeta$ required to match observational data is not known {\em a
priori}, in all of the work presented in this paper we fix
$\gamma_{\rm max} = 10^{5}$, which is also consistent with the maximum
$\gamma$ at which relativistic electrons are frozen into the
post-shock flow.

\subsection{NT emission and absorption}
With a power-law energy distribution, the magneto-bremsstrahlung
emission can be calculated analytically, as in Paper~I. However, given
an arbitrary energy distribution of non-thermal electrons,
calculation of the magneto-bremsstrahlung emission from the downstream
flow requires a numerical integration of the magneto-bremsstrahlung
emissivity over the full energy range of the relativistic electron
energy spectrum. 

\subsubsection{Synchrotron emission}
\label{sec:sync}
In the synchrotron limit, the emission arising from a
single relativistic electron with Lorentz factor $\gamma$ is

\begin{equation}
\label{eq:sync_single}
P(\nu) = \frac{\sqrt{3}q^{3}\beta^2B{\rm sin}\alpha}
{m_{\rm e}c^{2}}F(\nu/\nu_c),
\end{equation}

\noindent where $\alpha$ is the pitch angle of the particle relative
to the direction of the B-field, and $F(\nu/\nu_c)$ is a dimensionless
function describing the total power spectrum of the synchrotron
emission, with $\nu_c$ the frequency where the spectrum cuts off,
given by

\begin{equation}
\label{eq:nu_c}
\nu_c=\frac{3\gamma^2 q B {\rm sin}\alpha}{4 \pi m_{\rm e} c} = \frac{3}{2}\gamma^2 \nu_{\rm b}{\rm sin}\alpha. 
\end{equation}

\noindent $\nu_{\rm b} = \omega_{\rm b}/2\pi$, where 
$\omega_{\rm b} = qB/m_{\rm e}c$ is the cyclotron 
frequency of the electron. $\omega_{0} = \omega_{\rm b}/\gamma$ is the 
orbital or relativistic gyration frequency of the electron. 
Values for $F(\nu/\nu_c)$ are tabulated in
\citet{Ginzburg:1965}. Throughout the remainder of this paper we
assume that ${\rm sin}~\alpha = 1$. While this means that our calculations
will slightly overestimate the synchrotron emission for a given
$\zeta_{\rm B}$, a more detailed 
consideration is not warranted at the present time.
At the stagnation point in our ``standard model'', 
$\nu_{c} = 260$~MHz for $\gamma=100$,
and 26~GHz for $\gamma=10^{4}$.
 
The synchrotron emissivity formula (Eq.~\ref{eq:sync_single}) may be used even 
for mildly relativistic electrons (e.g., $\gamma < 2$), provided that 
$\chi \gtsimm 100$, where

\begin{equation}
\label{eq:chi}
\chi = \frac{\nu}{\nu_{\rm b}} = \frac{\omega}{\omega_{\rm b}} =  
\frac{\omega}{\gamma \omega_{0}}
\end{equation}

\noindent \citep[see Fig.~2 of][]{Mahadevan:1996}.
Since $\chi \propto \nu/B$, $\chi$ is small when
we have low frequencies and high B-fields (or, alternatively when the
binary separation is small, since $B \propto 1/D_{\rm sep}$). The lowest 
frequency we examine is 200~MHz. In our ``standard model''
(see Sec.~\ref{sec:standard_model}), $\nu_{\rm b} = 1.7 \times 10^{4}$~Hz
and $B = 5.9$~mG at the stagnation point of the WCR, which yields
$\chi = 1.2 \times 10^{4}$ when $\nu = 200$~MHz. The declining B-field
means that $\nu_{\rm b}$ declines with off-axis distance, which leads
to higher values of $\chi$. In the most extreme case which we consider
($D_{\rm sep} = 2 \times 10^{14}$~cm, $\zeta_{\rm B} = 10^{-2}$), $\chi > 100$
throughout the WCR, justifying our use of the synchrotron emissivity formula.

\subsubsection{Synchrotron self-absorption}
\label{sec:ssa}
Synchrotron self-absorption (SSA) or, more accurately at low $\gamma$,
magneto-bremsstrahlung self-absorption, is intimately linked to the
energy spectrum of the non-thermal particles and their
magneto-bremsstrahlung emissivity, and in Appendix~A we show how the
calculation of SSA is performed.
The SSA turnover frequency for a source subtending a solid angle
$\Omega$ (steradians) is given by \citep[for $p=2$;
\cf][]{Pacholczyk:1970}
\begin{equation}
\label{eq:nu_ssa}
\nu_{\rm SSA} = 46.3 \;B^{1/6} S_{\nu}^{1/3} \nu^{1/6} \Omega^{-1/3} 
\;{\rm Hz},
\end{equation}
where $B$ is in Gauss, and $S_{\nu}$ is the
synchrotron flux in mJy at any frequency $\nu$ (in Hz) where the source
is optically thin. For our ``standard'' CWB model (without IC cooling)
with $\zeta = 10^{-4}$ (see Sec.~\ref{sec:standard_model}), 
the magnetic field strength at the apex of the WCR is $5.9\;{\rm mG}$, 
$S_{\nu} = 0.026 \;{\rm mJy}$ at $\nu = 81.1 \;{\rm GHz}$, and 
$\Omega \sim 3 \times 10^{-15}\;{\rm ster}$ (the spatial extent of
synchrotron emission from the WCR is $\sim 60 \times 5$~mas), 
giving $\nu_{\rm SSA} \approx 20\;{\rm MHz}$\footnote{Note that the 
calculations in Paper~I greatly overestimate the strength of SSA (see, 
e.g., Fig.~4 in that paper). This error resulted from the use of Eq.~6.53
in \citet{Rybicki:1979} that takes $C$ from Eq.~6.20a. We had assumed it
was taken from Eq.~6.20b, since this is the case for Eq.~6.36.
This led to the erroneous conclusion in Paper~I that the SSA turnover
frequency and the characteristic cut-off frequency from the Razin effect
are comparable in our ``standard'' CWB model when $\zeta = 10^{-3}$. We
now find that this is the case when $\zeta \sim 10^{-2}$ - see also
the discussion in Sec.~\ref{sec:var_dsep}.}.
For wide CWBs, SSA is negligible at GHz frequencies, but
may become important in closer systems.

\subsubsection{The Razin effect}
\label{sec:razin}
When relativistic charges are surrounded by a plasma (as opposed to
existing in a vacuum), the beaming effect that characterizes
synchrotron radiation is suppressed. As noted in Paper~I, the
refractive index of the medium reduces the Lorentz factor of the
electron to

\begin{equation}
\label{eq:razin_debeam}
\gamma' = \frac{\gamma}{\sqrt{1 + \gamma^{2}\nu_{0}^{2}/\nu^{2}}},
\end{equation}
 
\noindent where the plasma frequency 
$\nu_{0} = \sqrt{q^{2}n_{\rm e}/\pi m_{\rm e}}$.
In Paper~I, this effect was approximated as an exponential reduction
in flux at the characteristic cut-off frequency,

\begin{equation}
\nu_{\rm R} = 20 \frac{n_{\rm e}}{B}.
\label{eq:razin_freq}
\end{equation}

\noindent However, since we calculate the magneto-bremsstrahlung
emissivity for individual values of $\gamma$ in this work, we simply replace
$\gamma$ with $\gamma'$ in the calculations of the non-thermal emission
and absorption\footnote{We note that the Razin effect is slightly more subtle
than replacing $\gamma$ with $\gamma'$ in Eqs.~\ref{eq:sync_single} 
and~\ref{eq:nu_c}, since not all of the Lorentz factors in the function 
$F(\nu/\nu_{c})$ are related to the beaming effect - see, e.g., 
\cite{Rybicki:1979}, \cite{Ginzburg:1965}, or \cite{vanLoo:2004} for more
details. However, this has little impact on the results.}.

\subsection{Clumping}
\label{sec:clump}

There is a great deal of evidence that the winds of hot stars are
clumpy \cite[e.g.,][]{Moffat:1988,Robert:1994,Lepine:2000},
and the effect of inhomogeneous winds on the WCR is expected to be
sensitive to a number of variables
\cite[e.g.,][]{Cherepashchuk:1990,Lepine:1995}. Foremost amongst these
is whether the densities are low enough that the WCR and the shocked
clumps are adiabatic. This is indeed the case in the wide binary
systems which we consider in this paper\footnote{The interaction of a clumpy wind with a radiative WCR has been
discussed by \citet{Walder:2002}.  In this case the WCR may become relatively
narrow, though dense, and prone to instabilities, such as nonlinear
thin-shell instabilities \cite[e.g.,][]{Stevens:1992,Vishniac:1994}.
The fate of clumps in this situation is less certain.
It is possible that they will be destroyed when they encounter
the WCR, but since the post-shock gas is itself cold and dense, new
``clumps'' may be created as the WCR fragments under the action of
these instabilities. Of note is the fact that the growth rate of the
thin-shell instability depends on high
tangential velocities channelling material to the extremities of the
kinks \citep{Blondin:1996} - this channelling 
may be disrupted if the incident flows are clumpy, and the growth 
of this instability moderated to some extent, though a numerical calculation
would be needed for confirmation of this idea.}. In such a situation the
clumps are stripped of material by a combination of dynamical
processes (such as ablation) and evaporation (resulting from the
conduction of heat into their interiors) as they move within the WCR.

The destruction timescale of non-radiative clumps by the action of
dynamical instabilities is $t_{\rm d} = \epsilon t_{\rm cc}$
\citep{Klein:1994}, where $t_{\rm cc}$ is the cloud crushing
timescale, defined as $t_{\rm cc} = r_{\rm c}/v_{\rm s}$, $r_{\rm c}$
is the radius of the clump, and $v_{\rm s}$ is the shock velocity (for
the stationary shocks in CWBs, $v_{\rm s}$ is equal to the pre-shock
wind speed). $\epsilon \approx 3.5$ for a density ratio between the clump
and the inter-clump wind of order 10-100 \citep{Klein:1994}, though
larger density ratios may occur in intrinsic wind shocks.
A clump which passes through a shock experiences less
deceleration than the inter-clump material, so a lower limit on the
timescale for the clump to reach the contact discontinuity is $t_{\rm
cd} = \Delta r/v_{\rm s}$, where $\Delta r$ is the distance from the
shock to the CD. Clumps will be destroyed before they reach the CD
when
\begin{equation}
\label{eq:clump_dest}
t_{\rm d}/t_{\rm cd} = \epsilon t_{\rm cc}/t_{\rm cd} = 
\epsilon r_{\rm c}/\Delta r < 1.
\end{equation}
There are indications that $r_{\rm c} \sim 1 R_{*}$ at distances of
$\sim 20 R_{*}$ from the star \citep{Lepine:2000,Rodrigues:2000}.  Such
clumps cannot expand faster than their sound speed, unless they behave
isothermally, in which case they may expand a few times faster.  The
wind temperature of hot stars is $\sim 1-10$~kK, and the sound speed
of the clump is $\sim 10\kmps$. In \object{\wr147}, the flow time of
the WR wind to the stagnation point is $\sim 10^{8}$s, so the clumps
will have expanded to radii of $\sim 1000\;\Rsol$.  In our model of
\object{\wr147} in Sec.~\ref{sec:wr147}, $\Delta r \approx 0.05 D_{\rm sep}$,
when measured along the line of centres of the stars. We thus
determine that $t_{\rm d}/t_{\rm cd} \sim 0.5$, which implies that the
clumps are rapidly destroyed. Material originating in the clumps is
mixed in with the post-shock gas and the high sound speed of the hot
WCR works to minimize the density and pressure gradients existing in
the vicinity of the clump. The post-shock flow should rapidly become
smooth, and the global position of the shocks and the CD should be
affected very little by the clumpy nature of the winds.

If $t_{\rm d}/t_{\rm cd} >> 1$, clumps could in theory pass completely
through the WCR and into the unshocked wind of the companion star,
where they would be surrounded by a bowshock and subject to a high
level of ablation, before being destroyed. The possibility of clumps
passing through the WCR and interacting with the photosphere of the
companion star is considered by \cite{Marchenko:2002}.

Another consideration is whether new clumps are created when gas in a
WCR which is largely adiabatic eventually cools. We make the general 
statement that the growth of density perturbations within the WCR 
will depend on their size relative to the local cooling length of the
surrounding gas \citep[see][]{Blondin:1989}, and note that any non-thermal
component of the pressure (e.g., from relativistic particles) lessens
the growth rate of the local thermal instability by decreasing the
dependence of the total pressure on the gas pressure
\citep[e.g.,][]{Wagner:2005}. Since we know that there are non-thermal
particles within the WCR we assume that the cold component of the
WCR is smooth. 

To summarize, we can distinguish the following three regions where
there may or may not be clumps: stellar winds; hot WCR; cold WCR. We
have demonstrated that clumps are rapidly destroyed in systems where
the WCR is largely adiabatic, and have noted that clumps are less
likely to be created within the WCR when there is supporting pressure
from non-thermal particles.  Therefore, we assume that all parts of
the WCR are smooth, leaving only the nature of the stellar winds to be
specified. Since the predominant effect of clumping is to introduce a 
scaling factor to the thermal emission, in our ``standard'' model in 
Sec.~\ref{sec:standard_model} we assume that the stellar winds are smooth.
However, our models for \object{\wr147} in Sec.~\ref{sec:wr147} adopt
clumpy winds in order to match the observed free-free emission with
specific mass-loss rates.

\subsection{Summary of the calculations}
\label{sec:sum_em}
Our calculation of the observed radio emission from CWB systems is
based on the density and temperature distribution obtained from a
hydrodynamical simulation of the stellar winds and their
collision. Both the stellar winds and the WCR contribute thermal
emission and absorption.  Free-free emission and absorption from
thermal electrons is calculated as noted in Paper~I, and clumping may
or may not be included in the model. In addition, the WCR has emission
and absorption components arising from a distribution of non-thermal
electrons accelerated at the shocks and advected with the downstream
flow.

To describe the non-thermal electron energy distribution we use 100,
logarithmically spaced, energy bins, extending from $\gamma = 1$ to
$\gamma_{\rm max}$, and which we specify as a power-law at the shocks.
The pre-cooled normalization of this energy spectrum in each cell
within the WCR is obtained by setting the total non-thermal electron
energy density to a specified fraction of the thermal energy density
(see Eq.~\ref{eq:nt_en_dens}).  Scalars in the hydrodynamical code are
then used to apply IC and ionic cooling, and a ``cooled'' non-thermal
electron energy distribution is obtained in each hydrodynamical cell.
As a result, the energy distribution in each cell within the WCR
shows some degree of evolution away from its initial power-law (see, e.g.,
Fig.~\ref{fig:standard_espec}a). In the downstream flow, the energy loss
from the non-thermal electrons means that $U_{\rm rel} \leq \zeta U_{\rm th}$.
Since the magnetic energy density in each hydrodynamical cell remains tied
to $U_{\rm th}$, the magnetic field and non-thermal electron energy
densities gradually diverge, with $U_{\rm B}$
exceeding $U_{\rm rel}$ in the downstream flow.
Electrons which cool to $\gamma=1$ are assumed to return to the thermal pool.

With the non-thermal electron energy distribution and the B-field
specified in every hydrodynamical cell within the WCR, the non-thermal
emission and absorption may be calculated. If the Razin effect is
specified, the de-beaming is first calculated and the effective
$\gamma$ of each energy bin is reduced according to
Eq.~\ref{eq:razin_debeam}.  The magneto-bremsstrahlung emissivity per
electron is then calculated using Eq.~\ref{eq:sync_single}.
The magneto-bremsstrahlung emissivity per unit volume is obtained from
multiplying the emissivity per electron by the number density of
non-thermal electrons. Integration over $\gamma$ then gives the 
total non-thermal emission per unit volume. Synchrotron self-absorption
is calculated in a similar fashion. Finally, the radiative transfer
equation is solved to generate synthetic images and spectra.

\begin{figure}[t]
\vspace{6.5cm}
\includegraphics{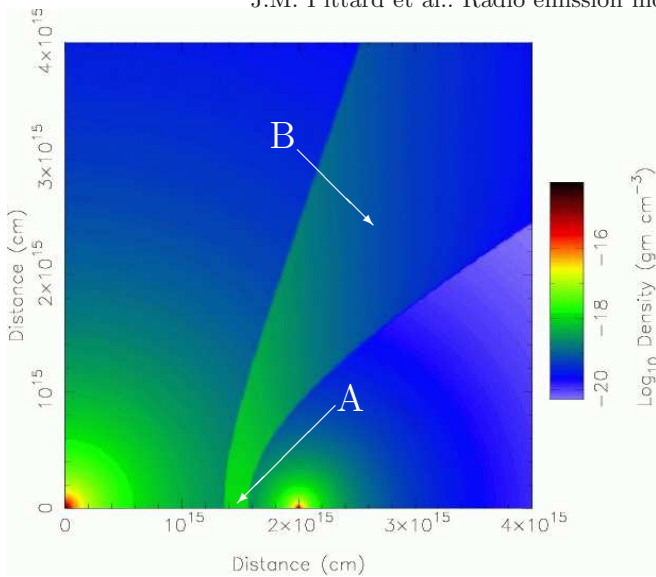}
\begin{center}
\setlength{\unitlength}{1mm}
\begin{picture}(110,1)(0,0)
\linethickness{1mm}
\color{white} 
\put(40,54){\vector(1,-1){10}}
\put(38,56){\makebox(0,0){\Large B}}
\put(45,20){\vector(-1,-1){13}}
\put(47,21){\makebox(0,0){\Large A}}
\color{black}
\end{picture}
\end{center}
\caption[]{A density plot of the standard model (showing only the 
positive z-region). The WR star is at (0,0). To aid subsequent 
discussion, two points in the post-shock flow are highlighted:
a position close to the stagnation point (position ``A''), and a position 
some distance off-axis in the shocked gas (position ``B'').}
\label{fig:standard_dens}
\end{figure}

\subsection{Some simplifying assumptions}
The simplifying assumptions mentioned in Sec.~2.6 of Paper~I apply
also to the work in this paper. An additional assumption which was not
discussed previously relates to the issue of thermal cooling of hot
plasma in the WCR. This is of relevance since the magnetic energy
density $U_{\rm B}$, and the energy density of relativistic electrons
$U_{\rm rel}$, are assumed to be related to the thermal energy density
$U_{\rm th}$. When the separation between the stars is sufficiently
small that significant cooling of the thermal plasma takes place
($\dot{E} \propto 1/D_{\rm sep}$), a two-fluid implementation of the post-shock
flow is ideally needed. 

Finally, we emphasize that in our current model the hydrodynamics of the
WCR is solved in the limit that a negligible amount of energy is
placed into non-thermal particles. In reality, efficient particle
acceleration may greatly modify the structure of the 
shocks bounding the WCR and the post-shock flow. Our work in
Sec.~\ref{sec:wr147} indicates that such changes may actually take
place.

\begin{figure*}[t]
\vspace{13.5cm}
\includegraphics{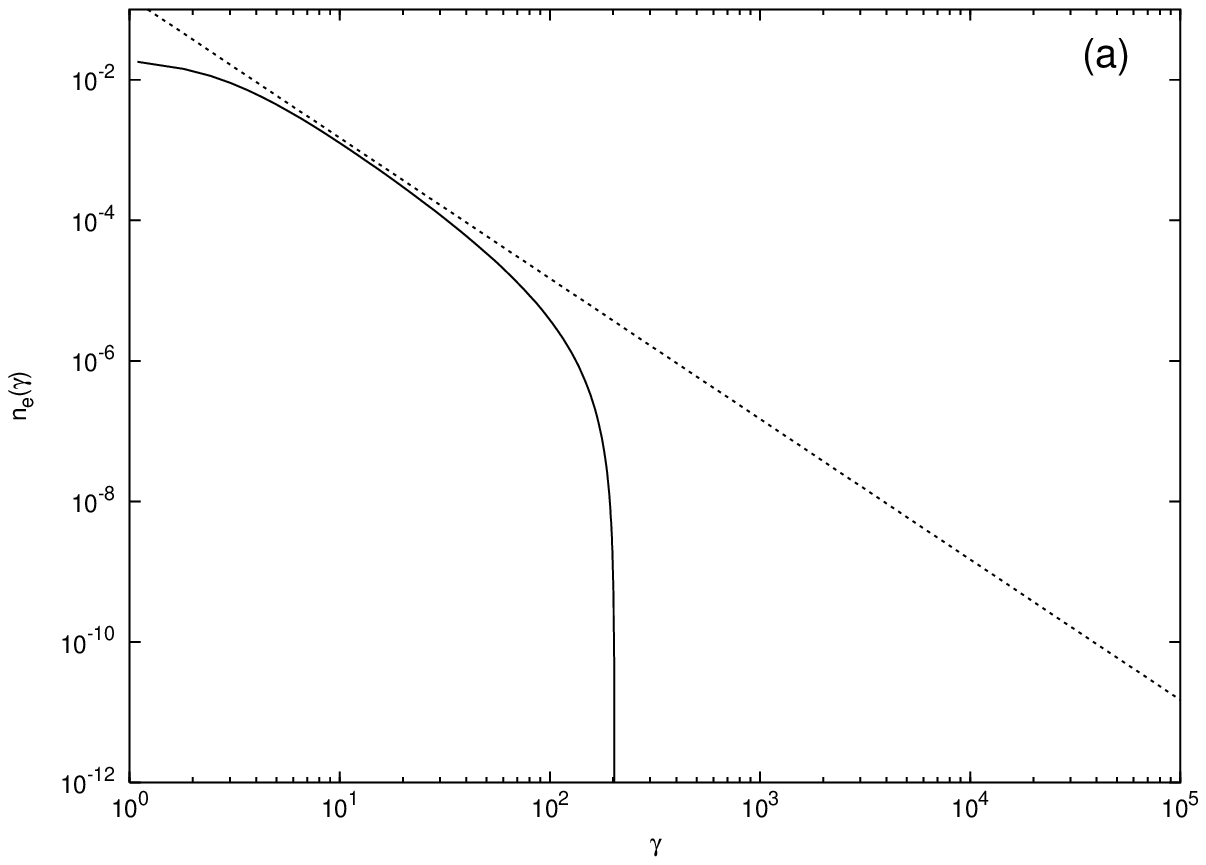}
\includegraphics{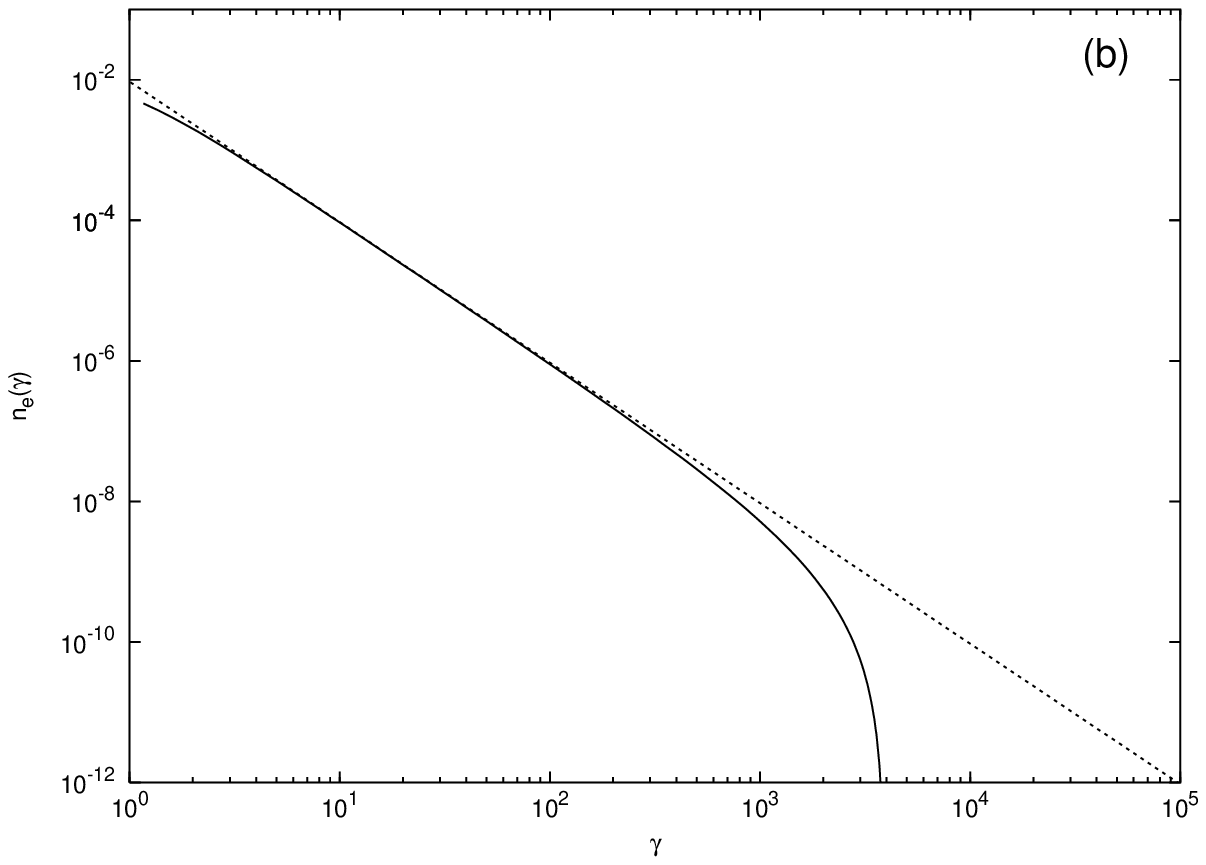}
\includegraphics{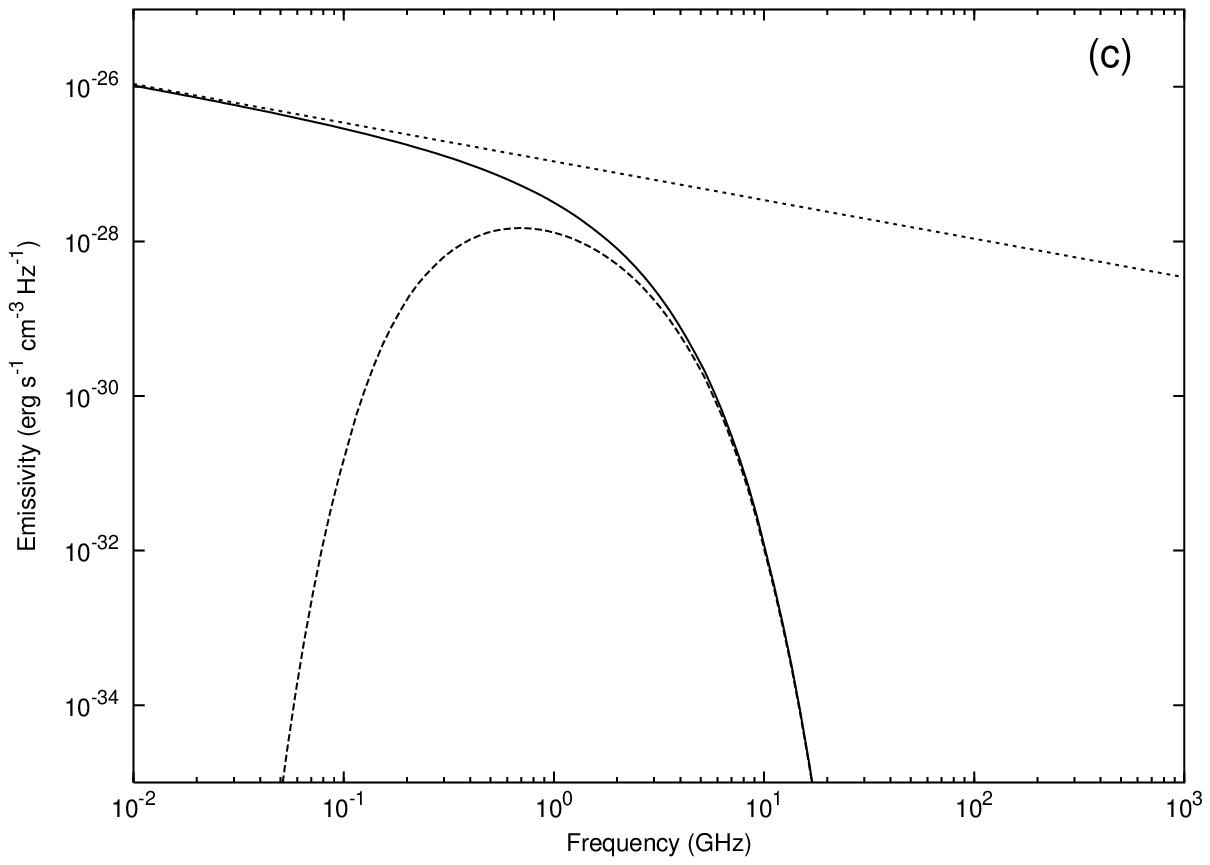}
\includegraphics{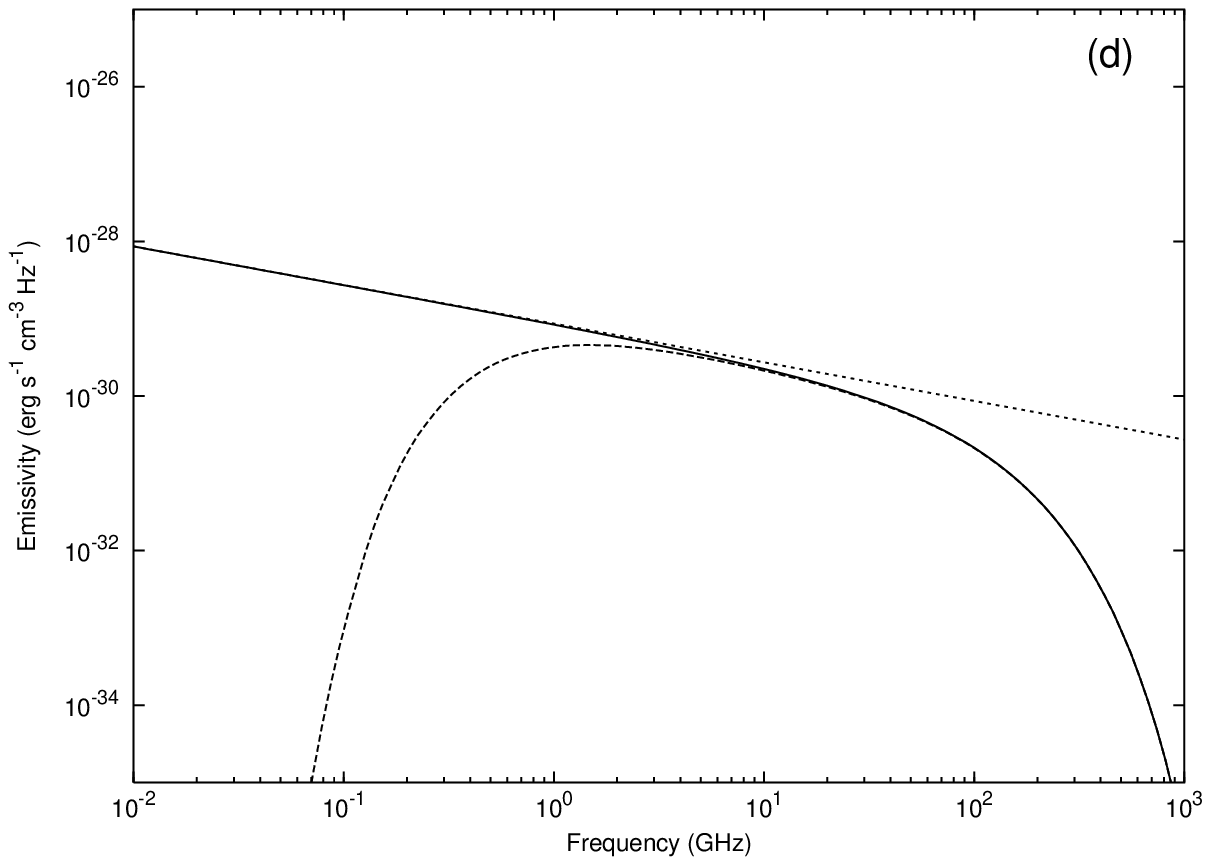}
\caption[]{The effect of IC and ionic cooling on the relativistic
electron energy spectrum (panels a and b) and on the resulting
synchrotron emission (panels c and d) for a position close to the
stagnation point (position ``A'', left panels) and some distance off-axis in
the shocked gas (position ``B'', right panels), as indicated in
Fig.~\ref{fig:standard_dens}.  In a) and b) the solid line shows the
$p = 2$ power-law spectrum assuming no cooling, while the dashed line
shows the effects of IC (high energy cut-off) and ionic cooling (low
energy roll-off). In c) and d) the magneto-bremsstrahlung emissivity
resulting from these energy distributions is shown.  
The low frequency cutoff can be influenced by a
number of processes. Here, the dashed curves show the Razin effect as
an example. The emissivity when IC and ionic cooling are or are not
included is shown as a solid and dotted curve respectively. The difference
between the solid and dotted curves indicates the improvements in this
paper relative to Paper~I.}
\begin{center}
\setlength{\unitlength}{1mm}
\end{center}
\label{fig:standard_espec}
\end{figure*}

\section{Parameter study}
\label{sec:param_study}

\subsection{A standard model}
\label{sec:standard_model}

As with our initial investigations in Paper~I, we examine the radio
emission from a ``standard'' CWB model with the following parameters:
$\Mdot_{\rm WR} = 2 \times 10^{-5}\;\Msolpyr$, $\Mdot_{\rm O} = 2
\times 10^{-6}\;\Msolpyr$, $v_{\infty,{\rm WR}} = v_{\infty,{\rm O}} =
2000\;\kmps$, $D_{\rm sep} = 2 \times 10^{15}\;\cm$. The wind momentum
ratio, $\eta = \Mdot_{\rm O} v_{\infty,{\rm O}}/\Mdot_{\rm WR}
v_{\infty,{\rm WR}} = 0.1$, and the distance of the stagnation point
from the WR and O star is respectively $r_{\rm WR} = 0.76D_{\rm sep}$
and $r_{\rm O} = 0.24D_{\rm sep}$. With such parameters the WCR is
largely adiabatic.  We assume solar abundances for the O star and
WC-type abundances for the WR star (mass fractions
$X=0,Y=0.75,Z=0.25$). Temperatures of $10,000$~K and an ionization
structure of ${\rm H^{+}}$, ${\rm He^{+}}$ and ${\rm CNO^{2+}}$ are
assumed for the unshocked stellar winds.  The wind temperatures are
lower than those used in Paper~I, but as the thermal flux from the
unshocked winds is only weakly dependent on the wind temperature
(through the gaunt factor), this change has only a small effect on the
free-free emission (see Sec.~\ref{sec:standard_spec_image}).  As
before, we adopt a distance of 1.0~kpc for our model system, and
assume that $\zeta=10^{-4}$.  As noted in Sec.~\ref{sec:ssa}, SSA is
negligible at GHz frequencies in the standard model. However, since $B
\propto \zeta^{1/2}$, and $S_{\nu} \propto \zeta^{7/4}$ in the
optically thin limit, we find that $\nu_{\rm SSA} \propto
\zeta^{2/3}$. The SSA turnover frequency is $\approx 0.2\;{\rm GHz}$
when $\zeta = 10^{-2}$.

Since IC cooling is now included in the models, appropriate
luminosities for the stars must also be defined. We set $L_{\rm WR} =
2 \times 10^{5} \Lsol$ and $L_{\rm O} = 5 \times 10^{5} \Lsol$ for
this first investigation which results in a photon energy density
$U_{\rm ph} = 2.3 \ergpcm3$ at the stagnation point\footnote{We estimate
$L_{\rm WR}$ and $L_{\rm O}$ from the following empirical
formulae. $L_{\rm WR}$ is calculated from the relationship ${\rm log}
\Mdot = 1.38 {\rm log} L - 12.1$ for Galactic WC stars
\citep{Crowther:2003}, while $L_{\rm O}$ is determined from ${\rm log}
\Mdot = 1.69 {\rm log} L - 15.4$ \citep{Howarth:1989}.}. For the
initial non-thermal energy spectrum at the shocks, we assume a
power-law energy distribution with $p=2$ and fix $\gamma_{\rm min} =
1$ and $\gamma_{\rm max} = 10^{5}$. This latter value is roughly 
the $\gamma$ at which the energy gain due to shock acceleration is
balanced by the energy loss due to IC cooling, for electrons near the
apex of the WCR - see Sec.~\ref{sec:gam_max}.

All of our 2D axisymmetric hydrodynamical calculations for the
standard model use a square grid with $0 \leq r \leq 4 \times 10^{15}
\cm$, $0 \leq z \leq 4 \times 10^{15} \cm$.  The z-axis is extended to
$-4 \times 10^{15} \cm$ with appropriate density and temperature
values for the unshocked WR wind before we process the grid through
our radiative transfer code. In our models, a viewing angle of 
$0^{\circ}$ implies lines-of-sight perpendicular to the axis of 
symmetry of the model.

\subsection{The thermal flux and $\tau=1$ surface of a hot star wind}
\label{sec:snu_rnu}

\begin{figure}[t]
\vspace{6.5cm}
\includegraphics{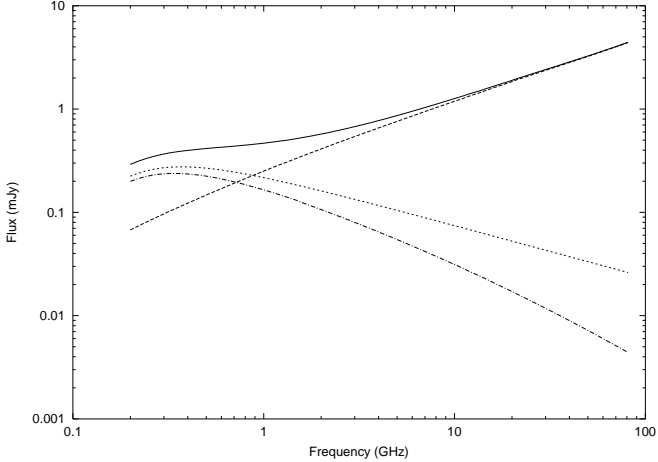}
\caption[]{Spectra from our standard model with a $0^\circ$ viewing angle
  - free-free flux (dashed), synchrotron flux (dotted), and
  total flux (solid).  We do not include IC cooling, the
  Razin effect or SSA. With IC 
  cooling included (dot-dash) a clear reduction in the amount of 
  synchrotron emission is seen.}
\label{fig:stan_spec1}
\end{figure}

To aid the interpretation of the results from our model, we calculate
the theoretical thermal flux from a single star, and the radius of
optical depth unity for a sightline from infinity.  As discussed by
\cite{Wright:1975}, the thermal radio flux for a radially symmetric,
isothermal wind moving at constant velocity is given by
\begin{equation}
\label{eq:sff}
S_{\rm ff} = 23.2 \left(\frac{\Mdot}{\mu_{\rm i}\vinfty}\right)^{4/3}
\gamma_{\rm e}^{2/3} g_{\rm ff}^{2/3} Z_{\rm i}^{4/3}\frac{\nu^{2/3}}{D^{2}} 
\, \mJy,
\end{equation}
where $\Mdot$ is in units of $\Msolpyr$, $\vinfty$ is in $\kmps$, the
distance to the system, $D$, is in kpc, and $\nu$ is in Hz. $Z_{\rm
i}$ is the rms charge of the ions, $\gamma_{\rm e}$ is the ratio of
electron to ion number density (i.e. $n_{\rm e} = \gamma_{\rm e}
n_{\rm i}$), $\mu_{\rm i}$ is the mean mass per ion (in a.m.u.), and
$g_{\rm ff}$ is the free-free Gaunt factor.  The radius at which a
given optical depth from infinity, $\tau$, occurs, is given by
\begin{equation}
\label{eq:rtau}
R_{\tau} = 1.75 \times 10^{28} \frac{\gamma_{\rm e}^{1/3} g_{\rm ff}^{1/3} 
Z_{\rm i}^{2/3} }{T^{1/2} \tau^{1/3}} \left(\frac{\Mdot}{\mu_{\rm i} 
\vinfty \nu}\right)^{2/3} \cm,
\end{equation}
where $T$ is in K. $R_{0.242}$ is equivalent to the
``characteristic radius'' defined by \cite{Wright:1975}.

If the wind is clumped (see also discussion in Sec.~\ref{sec:clump})
the thermal emission from the star mimics an increased mass-loss rate,
$\Mdot = \Mdot_{\rm actual}/\sqrt{f}$, where $f$ is the volume filling
factor of the clumps \citep{Lamers:1984}.  Values of the rms charge of
the ions ($Z_{\rm i}$), the ratio of electron to ion number density
($\gamma_{\rm e}$), and the mean mass per ion ($\mu_{\rm i}$) for the
standard model are 1.12, 1.09 and 4.88 for the WR star, and 1.00, 1.00
and 1.29 for the O star. Values of $S_{\rm ff}$, and $R_{\tau=1}=R_1$
are given in Table~\ref{tab:theory_single}.

\begin{table}[t]
\begin{center}
\caption[]{Theoretical values of the thermal radio flux, $S_{\rm ff}$,
and the radius of the $\tau=1$ surface, $R_{1}$, for the winds in our
standard model. A distance of $1\kpc$ is assumed, and the winds are
assumed to be smooth.}
\label{tab:theory_single}
\begin{tabular}{llll}
\hline
\hline
Parameter  & $\nu$ (GHz) & WR & O \\
\hline
$S_{\rm ff}$ (mJy) & 1.6 & 0.322 & 0.072 \\
             & 5.0 & 0.637 & 0.143 \\
             & 15.0 & 1.216 & 0.274 \\
$R_{1}$ ($10^{14}\cm$) & 1.6 & 4.07 & 1.93 \\
             & 5.0 & 1.83 & 0.87 \\
             & 15.0 & 0.84 & 0.40 \\
\hline
\end{tabular}
\end{center}
\end{table}

\begin{figure*}[ht]
\vspace{13.5cm}
\includegraphics{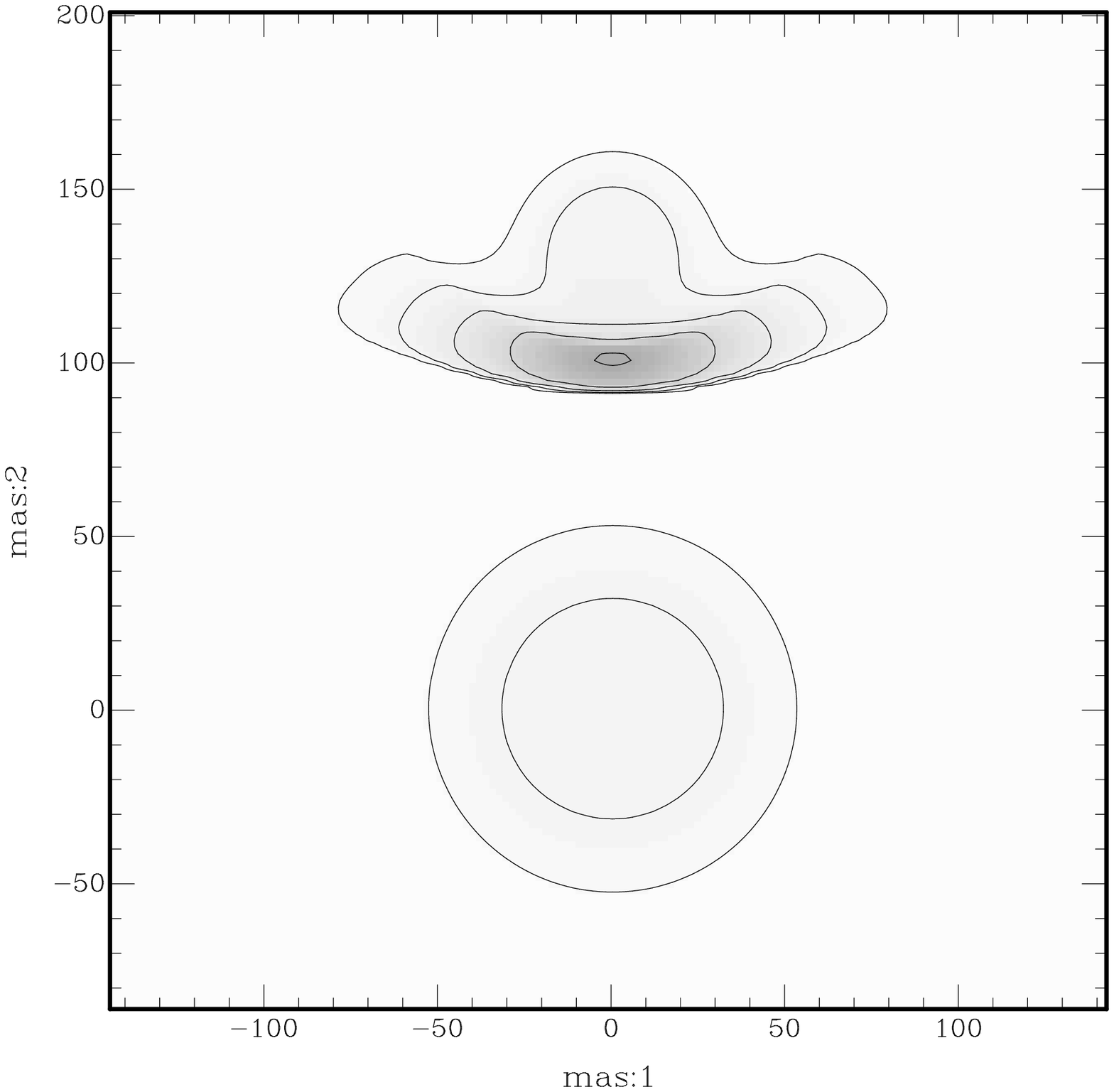}
\includegraphics{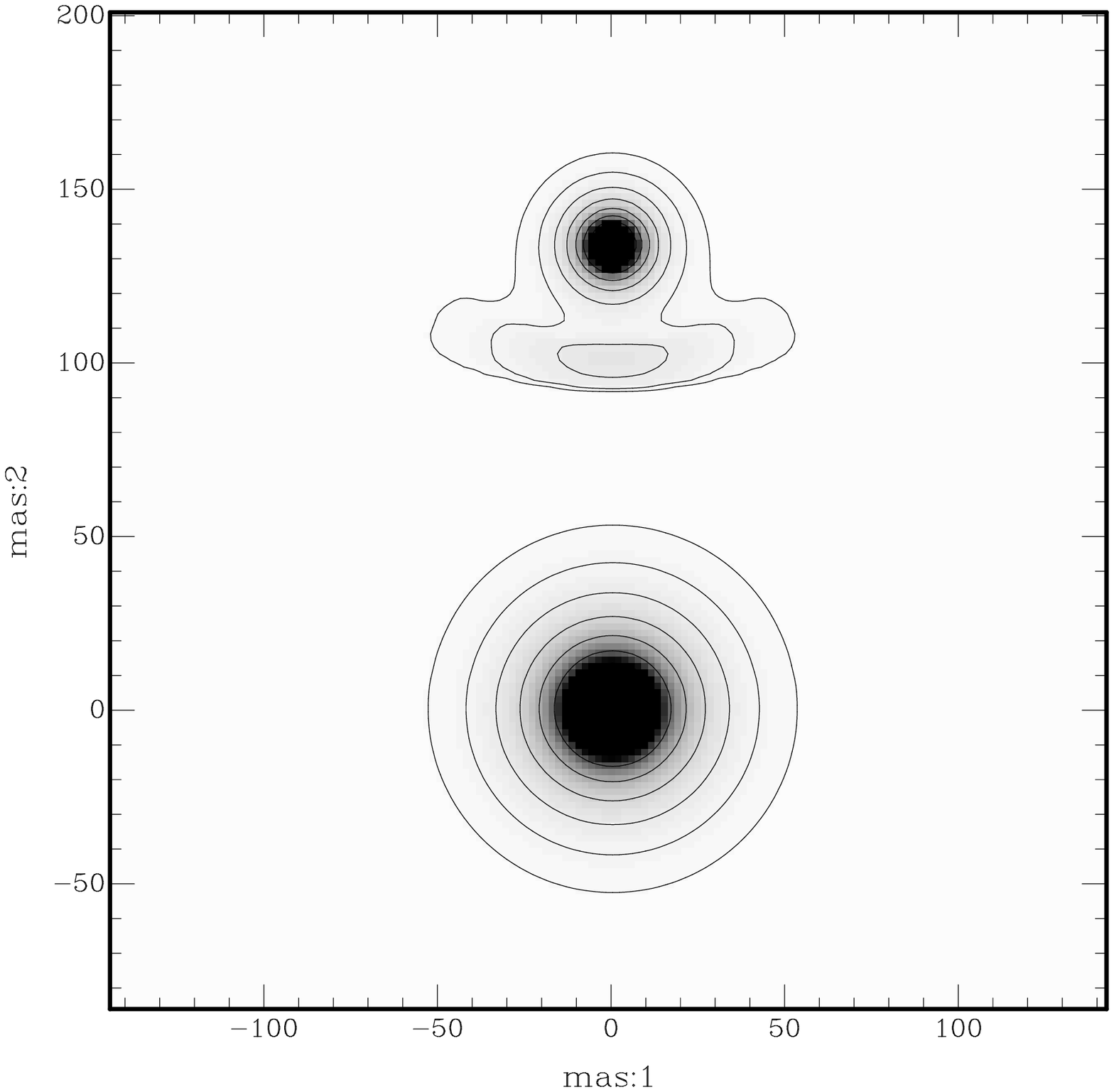}
\includegraphics{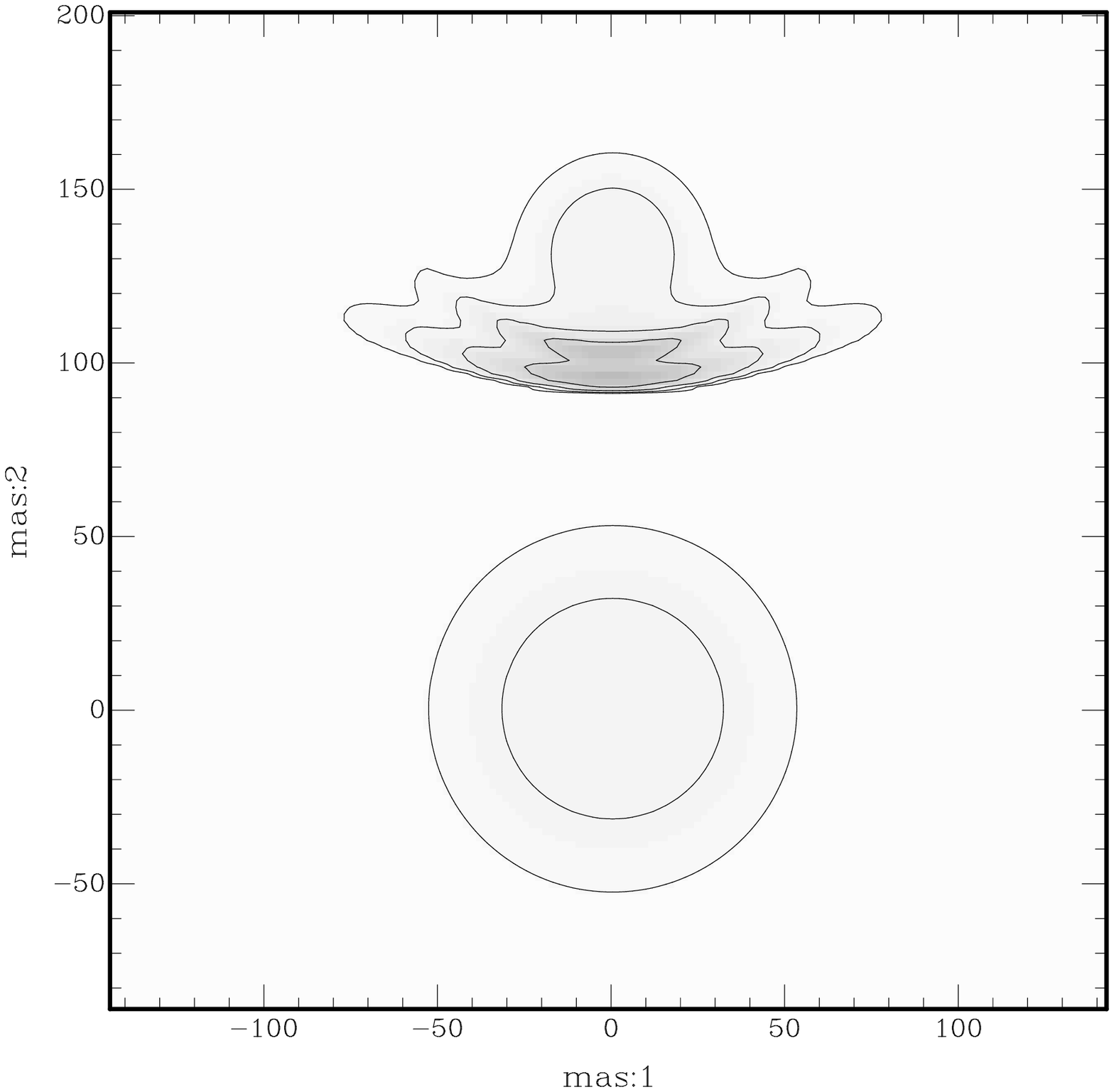}
\includegraphics{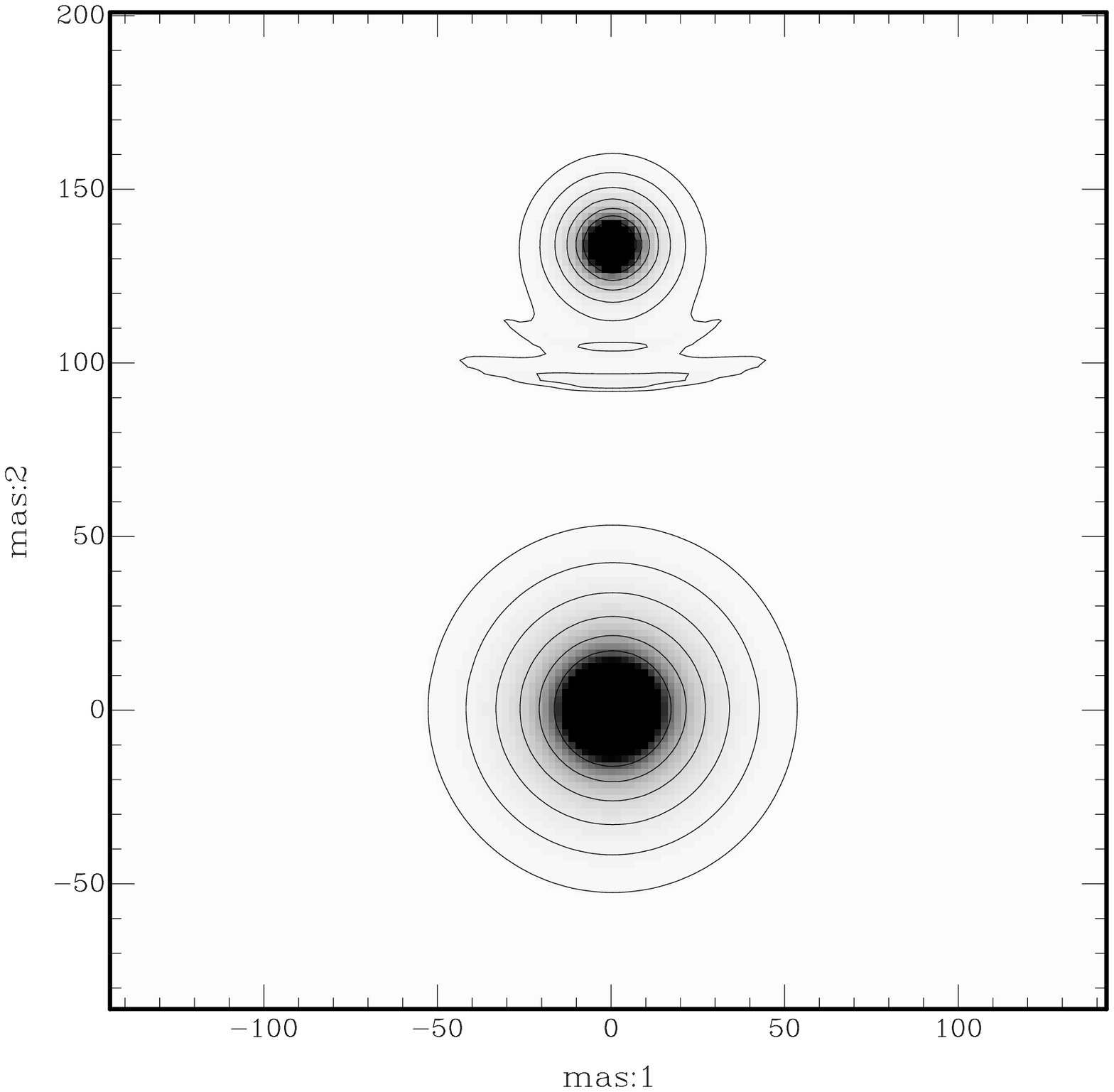}
\caption[]{The impact of IC cooling on the intensity distributions of our 
  standard model, for a viewing angle of $0^\circ$ and at
  1.6~GHz (top) and 22~GHz (bottom). Neither the Razin 
  effect nor SSA are included in this
  calculation. The images on the left do not include IC cooling, while
  those on the right do. Each image has the same intensity scale and
  contours.}
\label{fig:stan_image1}
\end{figure*}

\subsection{The salient features of the standard model}
\label{sec:salient}
Figs.~\ref{fig:standard_dens} and~\ref{fig:standard_espec}
illustrate the salient features of our modelling. The density
distribution of the unshocked stellar winds and WCR in our standard
model is shown in Fig.~\ref{fig:standard_dens}. Two locations in the
post-shock flow are highlighted, one near the stagnation point (``A''), and
another (``B'') at an off-axis position. 
Comparison of Figs.~\ref{fig:standard_espec}a and
\ref{fig:standard_espec}b yields several points worth noting.  First,
because the non-thermal electron energy density at ``A'' is greater
than that at ``B'', the ``initial'' energy spectrum (dotted line) has
a higher normalization in Fig.~\ref{fig:standard_espec}a.  Second, the
effects of IC and ionic cooling are more significant at ``A'' (solid
lines). IC cooling is high for flow close to the stars since the
intensity of the radiation field is high, and near the stagnation
point the flow speed is also low, so the exposure time is longer.  In
the standard model, IC cooling is responsible for the decline in
$\gamma_{\rm max}$ from $10^{5}$ (the assumed initial value) to
approximately $200$ for flow near the stagnation point
(Fig.~\ref{fig:standard_espec}a), while for the off-axis position this
decline stops at $\approx 4000$ (Fig.~\ref{fig:standard_espec}b). The
turn down in the electron spectrum at low $\gamma$ seen in
Figs.~\ref{fig:standard_espec}a and~\ref{fig:standard_espec}b is due
to ionic cooling.  This effect is again strongest for flow near the
stagnation point due to the higher ionic density and slower flow speeds
which exist there.

The magneto-bremsstrahlung emission resulting from the ``cooled''
relativistic energy spectra in Figs.~\ref{fig:standard_espec}a
and~\ref{fig:standard_espec}b are shown by the solid lines in
Figs.~\ref{fig:standard_espec}c and \ref{fig:standard_espec}d
respectively. The higher energy electrons that reside at position
``B'' produce magneto-bremsstrahlung emission extending to higher
frequencies ($\sim 10^{3}$~GHz, versus $\sim 10$~GHz at position
``A''; see also Eqs.~\ref{eq:sync_single} and~\ref{eq:nu_c}). Thus,
flow positions subject to high IC cooling produce
magneto-bremsstrahlung emission over a smaller range in frequency than
positions with less cooling.  With the Razin effect included, a
low-frequency cut-off to the magneto-bremsstrahlung spectra is seen as
a result of its de-beaming influence, as shown by the dashed curves in
Figs.~\ref{fig:standard_espec}c and~\ref{fig:standard_espec}d.  It is
also worth bearing in mind that the lower normalization of the
relativistic energy spectrum at ``B'' (Fig.~\ref{fig:standard_espec}b)
versus ``A'' (Fig.~\ref{fig:standard_espec}a) means that at 1~GHz the
emissivity per unit volume is over an order of magnitude higher at
position ``A''.

\subsection{Spectral and intensity distributions}
\label{sec:standard_spec_image}

The spectra and intensity distributions obtained from our standard
model are shown in Fig.~\ref{fig:stan_spec1} and the left column of
Fig.~\ref{fig:stan_image1}, respectively.  For simplicity, the only
absorption process included in Figs.~\ref{fig:stan_spec1} 
and~\ref{fig:stan_image1} is free-free absorption. 
Synchrotron emission dominates the total flux below 1~GHz.  The
turnover seen near 400~MHz is due to free-free absorption from the
unshocked stellar winds. Above 400~MHz, the synchrotron spectrum is a
power-law with a spectral index\footnote{The flux $S_\nu$ at frequency
$\nu$, $S_\nu \propto \nu^\alpha$, where $\alpha$ is the spectral
index.} $\alpha = -(p-1)/2 = -0.5$ for $p=2$ (see
Sec.~\ref{sec:modelling}).  The thermal spectrum is also a power-law
with a spectral index of $+0.6$, as expected.
The slight departure of the thermal emission from a power-law at low 
frequencies is due to the
finite size of our hydrodynamical grid. The thermal flux is
essentially the sum of the theoretical flux from each wind (\cf
Table~\ref{tab:theory_single}). However, this is not always the case
(see Sec.~\ref{sec:ff_vs_dsep} and Figs.~\ref{fig:ff2}
and~\ref{fig:ff_vs_dsep}).

The effect of IC cooling on spectra from our standard model is shown
by the dot-dash line in Fig.~\ref{fig:stan_spec1}. Its effect is
greatest at high frequencies, as demonstrated by the marked decrease
of the synchrotron spectral component. However, while this effect is
most severe at high frequencies, IC cooling reduces the flux in the
{\em entire} 0.1-100 GHz frequency range.  It is interesting that we
do not see a break in the synchrotron spectrum, in spite of its clear
signature in Figs.~\ref{fig:standard_espec}c
and~\ref{fig:standard_espec}d. This is because the frequency of the
break in the synchrotron spectrum is spatially dependent (compare
Figs.~\ref{fig:standard_espec}c and~\ref{fig:standard_espec}d), and
once integrated over the entire WCR the frequency break is smeared out
and thus not visible in the final spectrum. A clearly defined break
frequency is also lacking in single star models where the synchrotron
emission arises from shocks intrinsic to the wind \citep{vanLoo:2005}.

Spatially, the effect of IC cooling is to reduce the brightness of the 
synchrotron emission from the WCR, but more so near the CD than near the 
shocks (see Fig.~\ref{fig:stan_image1}).

\begin{figure}[t]
\vspace{6.5cm}
\includegraphics{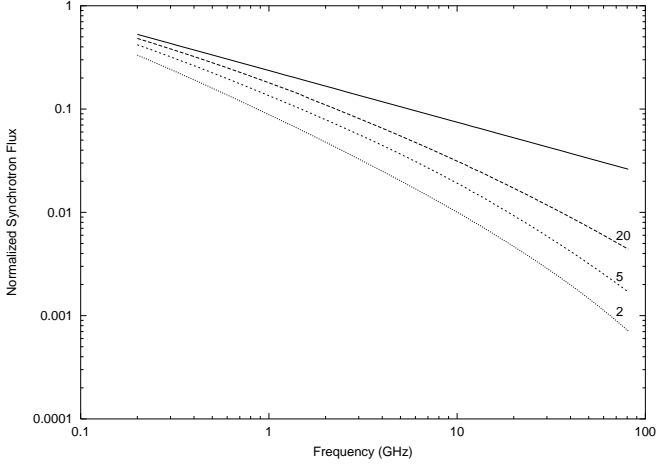}
\caption[]{The effect of separation on the IC cooled synchrotron
spectra with the standard model, a $0^\circ$ viewing angle and $\zeta
= 10^{-4}$, for $D_{\rm sep} = 2,5,$ and $20 \times 10^{14} \cm$. The
spectra are normalized so that the intrinsic emission spectra (i.e.
without free-free absorption, SSA, the Razin effect, or IC cooling)
are equal to the intrinsic spectrum at $D_{\rm sep} = 20 \times 10^{14} \cm$
(solid line).}
\label{fig:stan_spec2}
\end{figure}

\begin{figure}[]
\vspace{6.5cm} 
\includegraphics{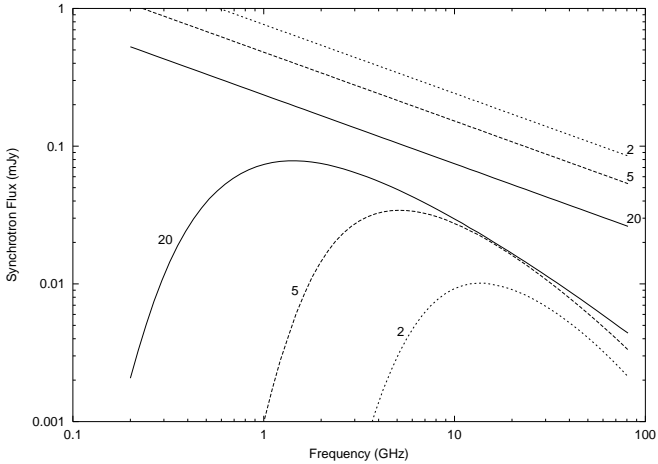}
\caption[]{The effect of separation on the synchrotron
spectra with the standard model, a $0^\circ$ viewing angle and $\zeta =
10^{-4}$, for $D_{\rm sep} = 2,5,$ and $20 \times
10^{14} \cm$. In each case the pre-IC cooled intrinsic synchrotron
spectrum and the observed IC-cooled synchrotron spectrum are shown.}
\label{fig:stan_spec5}
\end{figure}

\subsection{Impact of IC cooling and binary separation on the synchrotron 
emission}
\label{sec:var_dsep}
The dramatic effect of IC cooling on the resulting spectra and
intensity distributions should be even more pronounced when the
separation between the stars is smaller.  In Fig.~\ref{fig:stan_spec2}
we show the increasing effect of IC cooling on the synchrotron
emission as the stellar separation is decreased from $D_{\rm sep} = 2 \times
10^{15} \cm$ to $D_{\rm sep} = 2 \times 10^{14} \cm$. Since the intrinsic
synchrotron flux scales as $D_{\rm sep}^{-1/2}$, we have normalized the
pre-IC cooled spectrum of each model to that obtained at $D_{\rm sep} = 2
\times 10^{15} \cm$. This allows easier comparison of the divergence
between the intrinsic and IC cooled spectra as a function of $D_{\rm sep}$. The
clear trend is for the IC cooled synchrotron spectra to steepen with
decreasing $D_{\rm sep}$.

\begin{figure}[]
\vspace{18.0cm}
\includegraphics{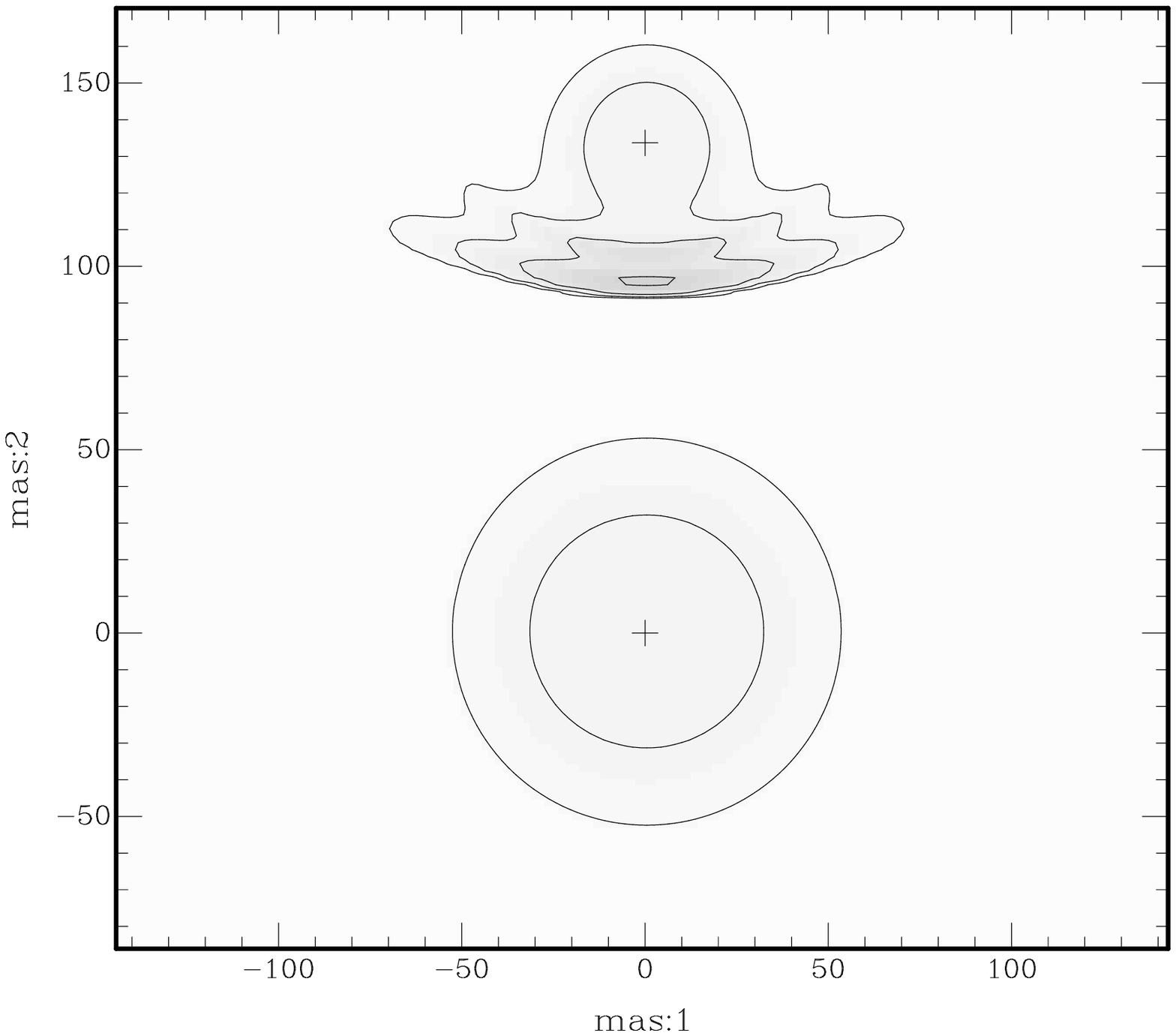}
\includegraphics{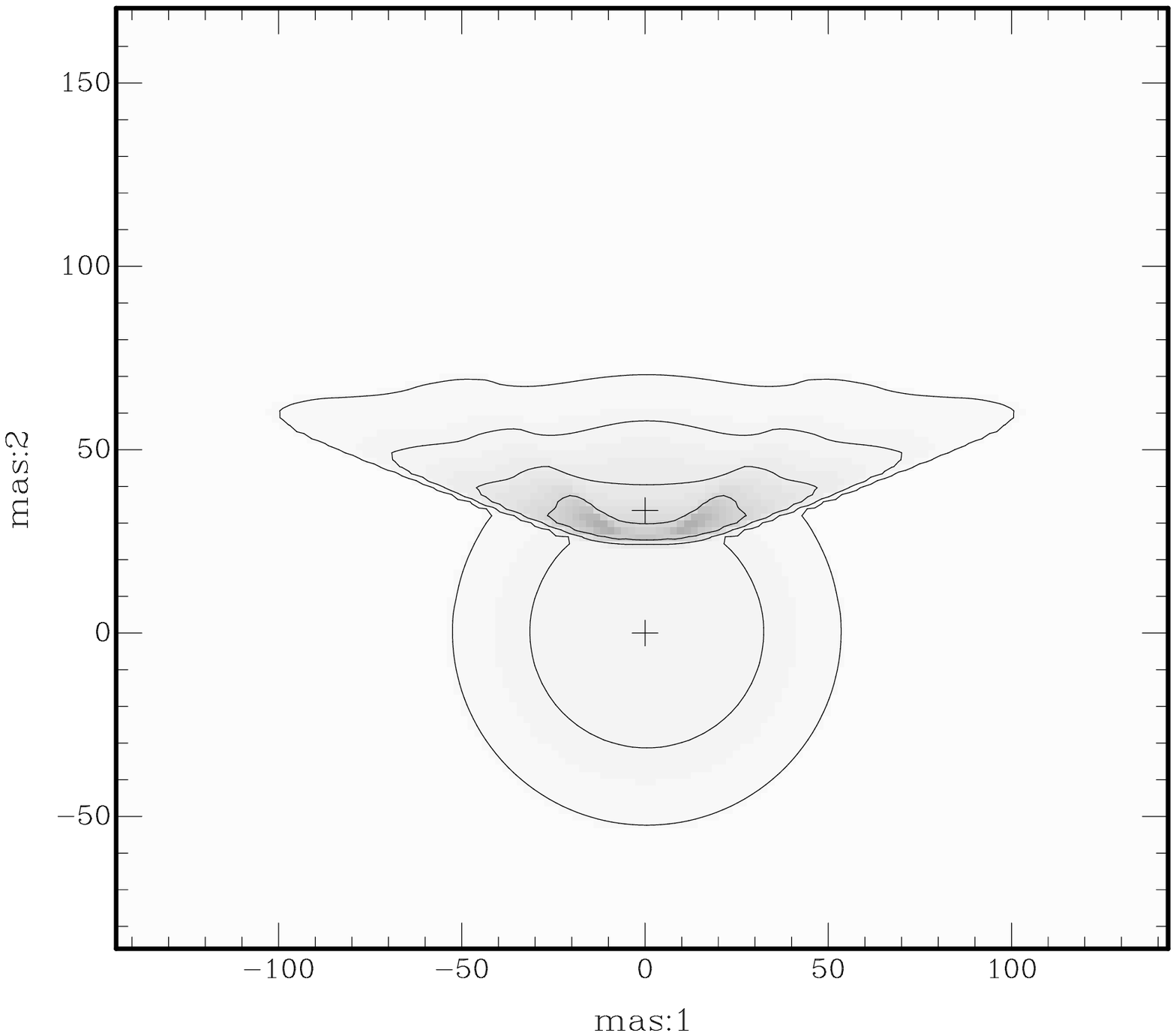}
\includegraphics{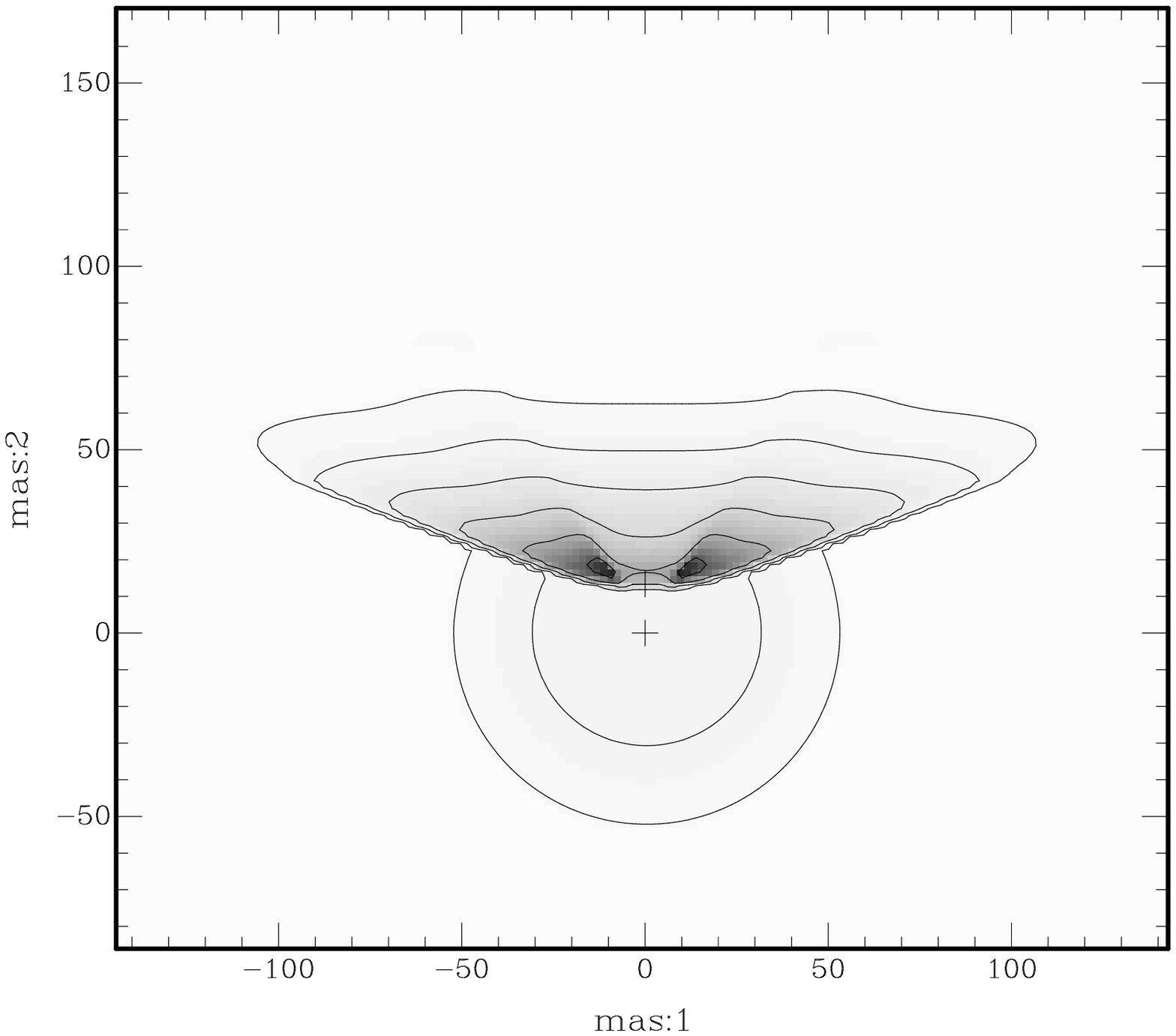}
\caption[]{Intensity distributions at 1.6 GHz and a $0^\circ$ viewing angle 
with a binary separation of 20 (top), 5 (middle), and $2
\times 10^{14}$~cm (bottom), corresponding to angular separations of
133, 33 and 13.3 mas respectively. The crosses denote the positions of
the stars, with the WR star located at (0,0). 
Each image has the same intensity scale and
contours. IC cooling, the Razin effect, SSA, and free-free absorption
are included, and $\zeta=10^{-4}$. Absorption by the WR wind is highest at
the apex of the WCR - as the binary separation decreases, the observed
intensity from the WCR reaches maxima either side of the symmetry
axis. The nature of the emission from the WCR also changes with binary
separation. In the top panel it is predominantly non-thermal, while
in the bottom panel it is completely thermal. This dramatic 
change is due to the Razin turnover frequency scaling as 
$\nu_{\rm R} \propto D_{\rm sep}^{-1}$.}
\label{fig:stan_image2}
\end{figure}

The spectra and intensity distributions from our full model (including 
IC cooling, the Razin effect, SSA and free-free
absorption) as a function of stellar separation are shown in
Figs.~\ref{fig:stan_spec5} and~\ref{fig:stan_image2}. While the
intrinsic synchrotron emission increases with decreasing $D_{\rm sep}$, 
the influence of IC cooling (at high frequencies) and free-free
absorption and the Razin effect (at low frequencies) means that the
{\em observed} synchrotron emission may decrease with decreasing
separation, which is consistent with the non-detection of non-thermal
emission from short period CWB systems. 

The low-frequency turnover in Fig.~\ref{fig:stan_spec5} is completely 
due to the Razin effect. Since $n_{\rm e} \propto \Mdot D_{\rm sep}^{-2}$,
and $B \propto \zeta^{1/2} \Mdot^{1/2} D_{\rm sep}^{-1}$, then $\nu_{\rm R}
\propto \zeta^{-1/2} \Mdot^{1/2} D_{\rm sep}^{-1}$, and we expect the
turnover to move to higher frequencies as the separation
decreases. This is indeed observed, and the turnover frequencies are
in excellent agreement with Eq.~\ref{eq:razin_freq}. In contrast, the
turnover frequency due to SSA (\cf Eq.~\ref{eq:nu_ssa}) scales as
$\nu_{\rm SSA} \propto \zeta^{2/3} \Mdot^{2/3} D_{\rm sep}^{-1}$,
since $S_{\nu} \propto \zeta^{7/4} \Mdot^{7/4} D_{\rm sep}^{-1/2}$ and
$\Omega \propto D_{\rm sep}^{2}$. Therefore the turnover frequencies
resulting from the Razin effect and SSA are both
$\propto 1/D_{\rm sep}$. If we ignore free-free absorption for the moment, we
find that the low-frequency turnover in our standard model is caused
by the Razin effect rather than SSA when $\zeta_{\rm B} = \zeta_{\rm
rel} \ltsimm 10^{-2}$ (departures from equipartition are examined in
the next subsection).  Since $\nu_{\rm SSA}/\nu_{\rm R} \propto
\zeta^{7/6} \Mdot^{1/6}$, the SSA turnover frequency may exceed
$\nu_{\rm R}$ when $\Mdot$ and particularly $\zeta$ are large.  The
turnover frequency due to free-free absorption depends on the viewing angle
into the system, and as it has a steeper
dependence on $D_{\rm sep}$ than either the Razin effect or SSA
($\nu_{\rm ff} \propto D_{\rm sep}^{-10/7}$ - see Paper~I), it is
increasingly likely that the low-frequency turnover will be caused by
free-free absorption as $D_{\rm sep}$ decreases.

At the high frequency end of the spectra in Fig.~\ref{fig:stan_spec5},
the difference between the observed and intrinsic synchrotron flux is
entirely due to IC cooling. With the B-field used in the
standard model, the emission at a given frequency arises from
electrons with fairly high values of $\gamma$ (\cf
Eq.~\ref{eq:nu_c}). Thus the influence of IC cooling is fairly
high, and this is reflected in the curvature in the synchrotron
spectra at high frequencies. As $D_{\rm sep}$ decreases, IC cooling becomes
increasingly important, and we find that high frequency
synchrotron emission is confined to very thin regions immediately
behind the shocks. The synchrotron emission behind the WR shock can be
appreciably higher than that behind the O shock, since the non-thermal
electrons are further removed from the O star in this case. 

Fig.~\ref{fig:stan_image2} shows that the maximum intensity of the
emission from the WCR may increase with decreasing binary
separation. However, the bottom panel of Fig.~\ref{fig:stan_spec5}
reveals that the synchrotron emission at 1.6~GHz when $D_{\rm sep} = 2 \times
10^{14}\cm$ is negligible, as it is greatly reduced by the Razin
effect. The observed emission from the WCR is therefore dominated by
free-free emission (the increase in the free-free emission from the
WCR with decreasing separation is investigated further in
Sec.~\ref{sec:ff_vs_dsep}).  The fact that the WCR emission at 1.6~GHz
changes from synchrotron dominated when $D_{\rm sep} = 2 \times 10^{15}\cm$, to
free-free dominated when $D_{\rm sep} = 2 \times 10^{14}\cm$, also explains why
the ``V''-shaped emission from the WCR seen at 
$D_{\rm sep} = 2 \times 10^{15}\cm$
disappears at $D_{\rm sep} = 2 \times 10^{14}\cm$ 
(see Fig.~\ref{fig:stan_image2}).

Fig.~\ref{fig:stan_image2} also shows that the emission from the WCR
is sometimes brightest to either side of the apex of the WCR, due to
high absorption.  This is sometimes seen in observations
\citep[e.g.,][]{Watson:2002,Dougherty:2005}, though it may also arise 
from time variability in the structure of the WCR.

\subsection{Departures from equipartition}
\label{sec:zeta_var}
So far, we have assumed that the energy density of the non-thermal
{\em electrons} and the B-field are equal. This assumption
was made for the sake of convenience, but it is quite probable that
this condition is not attained. For example, the B-field energy density
may be equal to the total non-thermal energy density from
electrons {\em and} ions. In fact, some astrophysical sources
are far from having $U_{\rm rel}=U_{\rm B}$ (e.g., in some very compact radio 
sources, \citet{Kellermann:1981} note that $U_{\rm rel}/U_{\rm B} >
10^{10}$!). We have therefore explored the effect
of varying the ratio $\zeta_{\rm B}/\zeta_{\rm rel}$. In essence, this
allows us to explore how changes in the B-field affect the emission.
The energy density of non-thermal ions has no effect on the radio emission.

\begin{figure}[t]
\psfig{figure=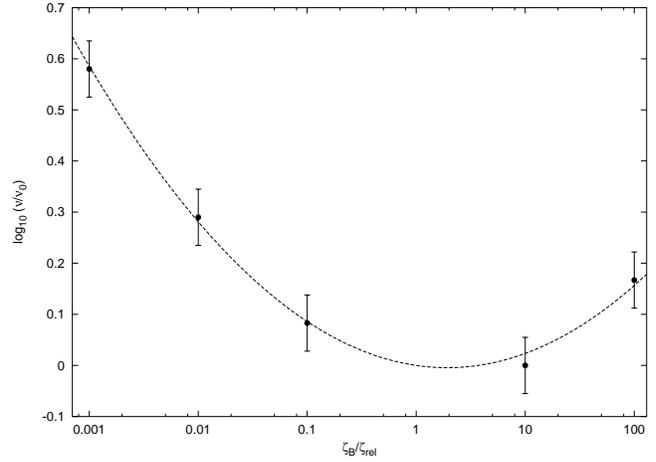,width=8.8cm}
\caption[]{The synchrotron turnover frequency (normalized to the
turnover frequency when $\zeta_{\rm B}=\zeta_{\rm rel}$) as a function of 
$\zeta_{\rm B}/\zeta_{\rm rel}$. Note that $\zeta_{\rm rel}$ was kept
constant in these calculations. We estimate the uncertainty of each
data point based on the finite frequency sampling in our calculations.
The curve is a least squares fit of a 
second order polynomial in ${\rm log}_{10} (\zeta_{\rm B}/\zeta_{\rm rel})$ 
to ${\rm log}_{10} (\nu/\nu_{\rm 0})$, and is forced to go through the 
position $(0.0,0.0)$. The coefficient's are $0.055$, $-0.031$ and 0.0.}
\label{fig:stan_spec7}
\end{figure}

When $\zeta_{\rm B}$ is kept constant
and $\zeta_{\rm rel}$ varied, the synchrotron spectrum simply scales in
flux as $S_{\nu} \propto \zeta_{\rm rel}$ (see Eq.~6 in Paper~I). However, when
$\zeta_{\rm rel}$ is kept constant and $\zeta_{\rm B}$ varied the
turnover frequency changes as well as the normalization, 
since $\nu_{\rm R} \propto 1/B \propto
1/\sqrt{\zeta_{\rm B}}$.  At high values of $\zeta_{\rm B}$ we 
expect SSA to succeed the Razin effect as
the dominant absorption process behind the low frequency turnover in
the synchrotron spectrum.  When this happens we expect the turnover
frequency to scale as $\nu_{\rm SSA} \propto B^{1/6}$ (see
Eq.~\ref{eq:nu_ssa}). Our calculations show that it is relatively easy
to achieve a turnover in the synchrotron spectrum at fairly high
frequencies through the Razin effect by reducing $\zeta_{\rm B}$, but
difficult to achieve turnovers through SSA even a little above that
obtained when $\zeta_{\rm B} = \zeta_{\rm rel}$ since the dependence
of $\nu_{\rm SSA}$ on $B$ is weak, and requires commensurately large
values of $B$. This is illustrated in
Fig.~\ref{fig:stan_spec7} where we plot the normalized turnover frequency 
as a function of $\zeta_{\rm B}/\zeta_{\rm rel}$. 

\subsection{The thermal flux and binary separation}
\label{sec:ff_vs_dsep}
The effect of binary separation on the {\em thermal} flux from the WCR has
not previously been considered.  
As noted by \citet{Stevens:1995}, the free-free emissivity from the
WCR is slightly enhanced compared to that from the
unshocked winds, at an equivalent distance from the system centre, since 
$\varepsilon_{\nu}^{\rm ff} \propto \rho^{2} T^{-1/2}
g_{\rm ff}$.  In the WCR the density is typically $4 \times$ higher
(\cf the jump conditions for a strong shock), the temperature is of
order $10^{2}-10^{4} \times$ higher, with perhaps a typical increase
of a factor of $10^{3}$, and the Gaunt factor is approximately double
that of the ambient gas. In contrast, the free-free absorption is
reduced by a factor of $\sim 10^{3}$, since
$\alpha_{\nu}^{\rm ff} \propto \rho^{2} T^{-3/2} g_{\rm ff}$. If the
unshocked winds are clumpy, and the gas in the WCR basically smooth
(see Sec.~\ref{sec:clump}), the
difference in absorption can be even larger.

\begin{figure}[t]
\psfig{figure=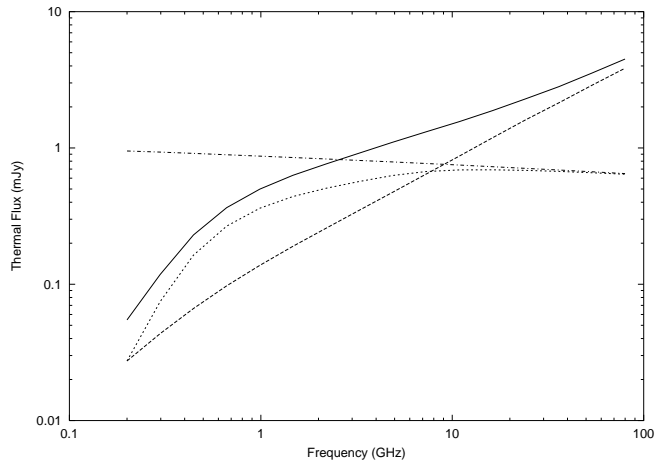,width=8.8cm}
\caption[]{The {\em thermal} emission from the standard model with 
a $0^\circ$ viewing angle and a binary separation of $2 \times 10^{14}\cm$. 
The thermal emission from the unshocked winds (dashed), the WCR (dotted), 
and the total thermal emission (solid) are shown. The intrinsic thermal 
emission from the WCR (before free-free absorption) is shown by the 
dot-dashed line. Synchrotron emission is not displayed.}
\label{fig:ff2}
\end{figure}

\begin{figure*}[ht]
\psfig{figure=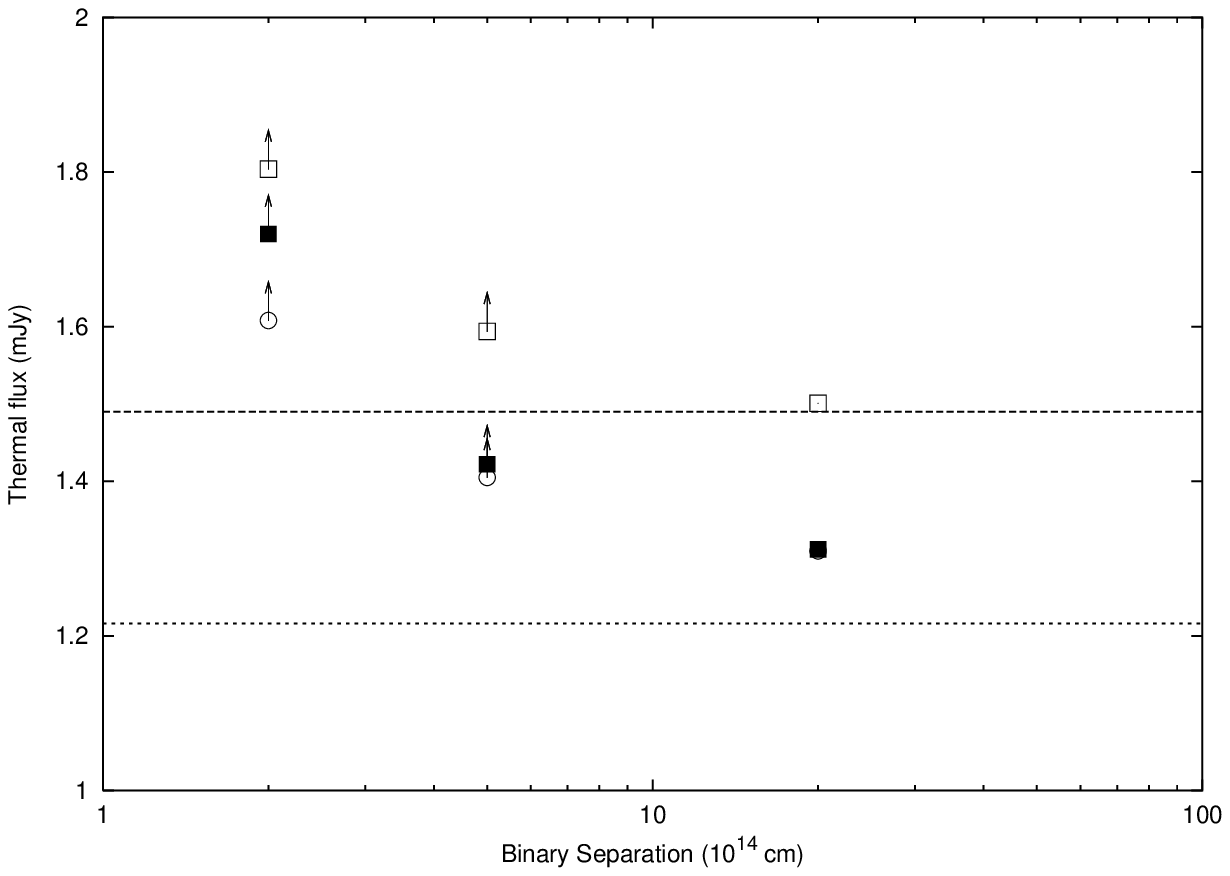,width=8.8cm}
\psfig{figure=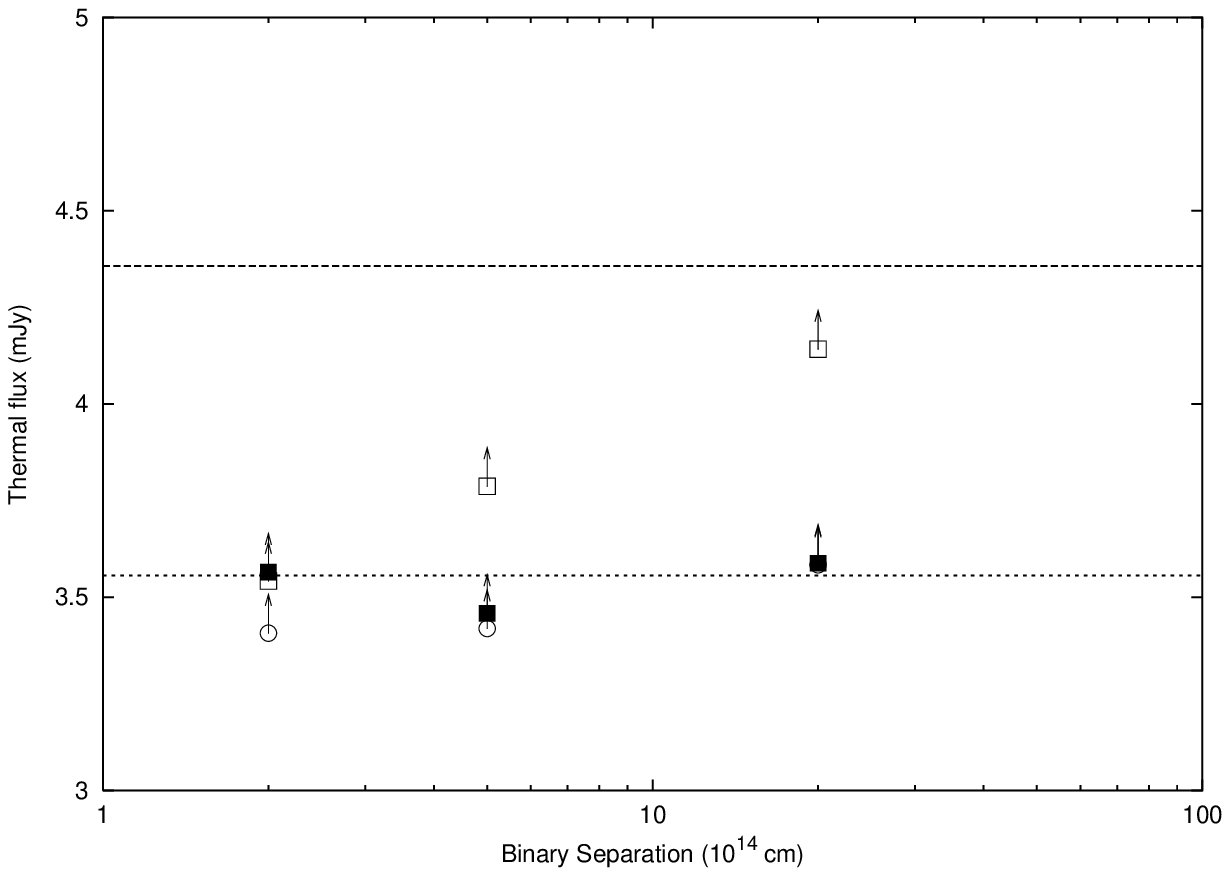,width=8.8cm}
\caption[]{a) The thermal flux from our standard model at 15~GHz and
various viewing angles as a function of the binary separation. The dotted
line shows the theoretical flux from the WR star,
while the dashed line shows the theoretical flux from the WR and O star
combined. The open circles are for a viewing angle of $-90^{\circ}$ 
(WR star in front), the open squares are for $0^{\circ}$, and the filled 
squares are for $90^{\circ}$ (O star in front). The points at 
$2 \times 10^{14}\;\cm$
and $5 \times 10^{14}\;\cm$ are plotted as lower limits, since not all
of the emission from the WCR is captured in the calculations. 
b) As a), but for clumpy winds with a filling factor of 0.2. In both
panels the open circle at $2 \times 10^{15}\;\cm$ is hidden underneath the 
filled square.}
\label{fig:ff_vs_dsep}
\end{figure*}

In Fig.~\ref{fig:ff2} we show the relative contribution to the total
thermal emission of the unshocked winds and the WCR.  Notably, the
high frequency thermal spectrum of the WCR has a spectral index which
is indicative of optically thin thermal emission, quite unlike that from the
unshocked stellar wind. The intrinsic thermal emission from the WCR is
strongly absorbed at low frequencies, mostly by free-free absorption
from the unshocked winds (i.e. from gas external to the WCR), with a
small contribution from absorption internal to the WCR.
The slight curvature in the unshocked wind emission at low frequencies
is a result of the finite size of the computational grid.

We find that the thermal emission from the WCR increases as the binary 
separation decreases, while the thermal emission from the unshocked winds 
remains
broadly the same. Since $\varepsilon_{\nu}^{\rm ff} \propto \rho^{2}
\propto D_{\rm sep}^{-4}$ and the volume of the WCR scales as $D_{\rm
sep}^{3}$, the thermal emission from the WCR scales as $D_{\rm
sep}^{-1}$. Since the emission from the WCR is optically thin, we 
see a substantial fraction of this emission at certain viewing angles, 
and at small separations the contribution of the WCR
to the total thermal flux can be significant. A dramatic consequence
of this is that a composite-like spectrum (Fig.~\ref{fig:ff2}), often
presumed to be evidence for non-thermal emission, can result entirely
from thermal processes! The spectral index between 5 and 15~GHz
of the total emission in Fig.~\ref{fig:ff2} is $\approx 0.4$, with the
result that the {\em free-free} emission from this model would be
classified as a ``composite'' spectrum if it was included in the
observational analysis of \cite{Dougherty:2000b}.  Nevertheless, we
believe that a non-thermal origin is the most likely explanation in
all of the systems noted in \cite{Dougherty:2000b} since these are 
typically wide systems.

In Fig.~\ref{fig:ff_vs_dsep}a we show how the total thermal flux varies
with binary separation and viewing angle at $\nu = 15$~GHz.  
At wide separations and a $0^{\circ}$ viewing angle, the total thermal 
flux from the system (the
sum of the thermal emission from the unshocked winds and from the WCR)
approaches the sum of the theoretical free-free flux from each stellar
wind.  This is because the WCR occurs well outside the $R_{1}$ radius
of emission of each wind (see Table~\ref{tab:theory_single}), and its
contribution to the emission is negligible since the intrinsic thermal
emission from the WCR scales as $D_{\rm sep}^{-1}$. In closer systems, the WCR
moves within the $R_1$ radius of each wind and removes a segment of
their emission. However, the resulting reduction in flux from the
unshocked winds is more than compensated for by the increase in
thermal emission from the WCR. Therefore, the total thermal emission
can considerably exceed that expected from the two stellar winds.  For
example, when $D_{\rm sep} = 2\times 10^{14}$cm, the thermal flux from the
standard model exceeds the total theoretical thermal flux by more than
20\%. The excess could be considerably higher at smaller separations
and lower frequencies. The effect is smaller at higher frequencies,
since $R_{1} \propto \nu^{-0.7}$.  Fig.~\ref{fig:ff_vs_dsep}a also
shows that the observed flux is reduced when one star is in front of
the other ($\pm 90^{\circ}$), as one would expect.

The nature and degree of clumping may also affect how the free-free
flux scales with binary separation. If the unshocked winds are clumpy,
but the WCR smooth due to rapid destruction of the clumps, the thermal
flux may actually decrease with decreasing separation, as shown in 
Fig.~\ref{fig:ff_vs_dsep}b. This is because as the stars move closer
together, the increased emission from the WCR is more than offset
by the decrease in the overall amount of clumpy material in the system.
Whether the thermal flux increases with decreasing binary separation
in a particular system depends on the nature and degree of clumping 
in that system.


\subsection{Synchrotron emission and wind-momentum ratio}
\label{sec:var_eta}
Free-free absorption by the surrounding stellar winds can have a large 
affect on the observable low frequency synchrotron emission. 
In previous models where the non-thermal emission is treated as 
arising from a point source, the free-free absorption by the stellar
winds is determined along a single line of sight. Since the spatial extent
of the WCR is ignored in such models, the low opacity within the WCR
is also disregarded. In our models, the spatial extent of the WCR means
that there are multiple lines-of-sight to the WCR. The amount of
free-free absorption that the synchrotron emission
suffers varies with viewing angle, but is also sensitive to the 
wind momentum ratio, $\eta$, since it is the combination of these 
parameters that determines how lines of sight into the system
are aligned with the opening angle of the WCR.

\begin{figure*}[t]
\psfig{figure=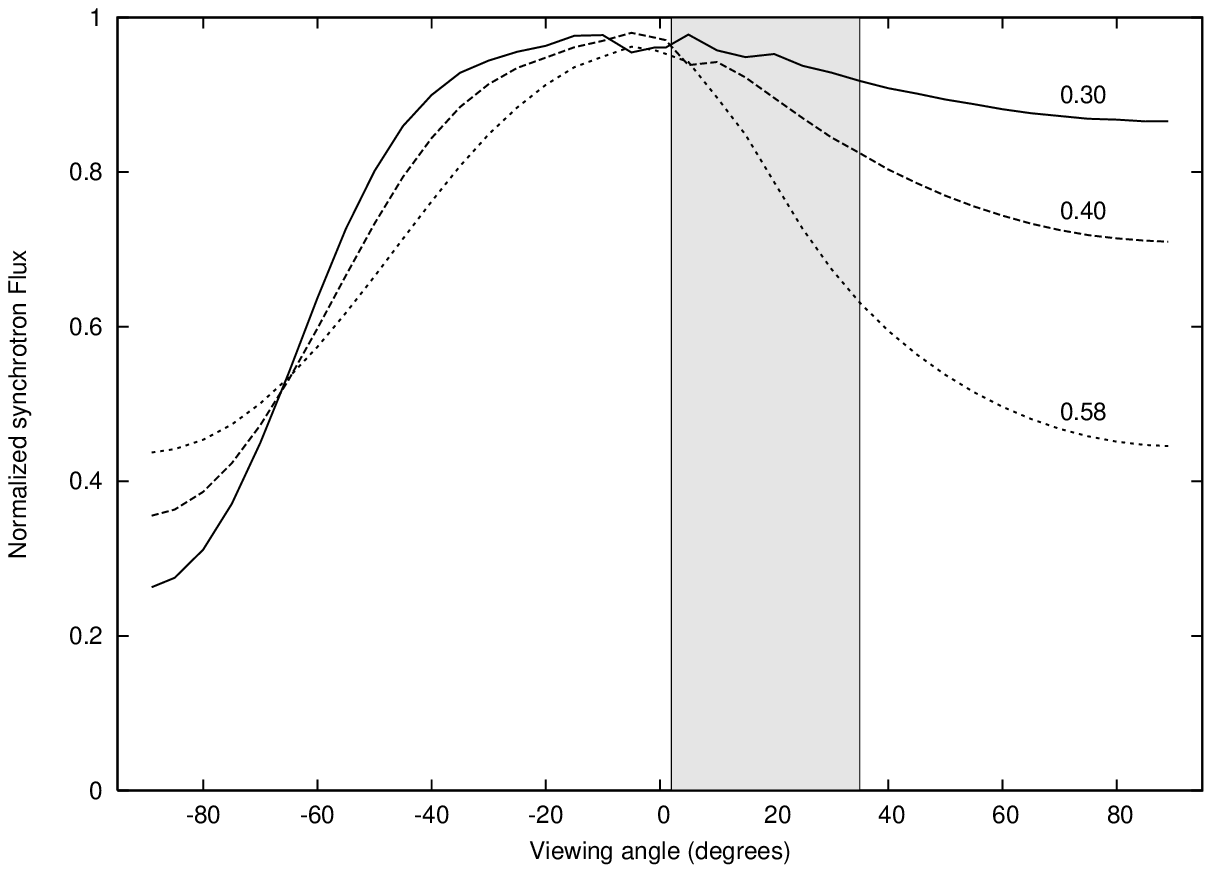,width=8.8cm}
\psfig{figure=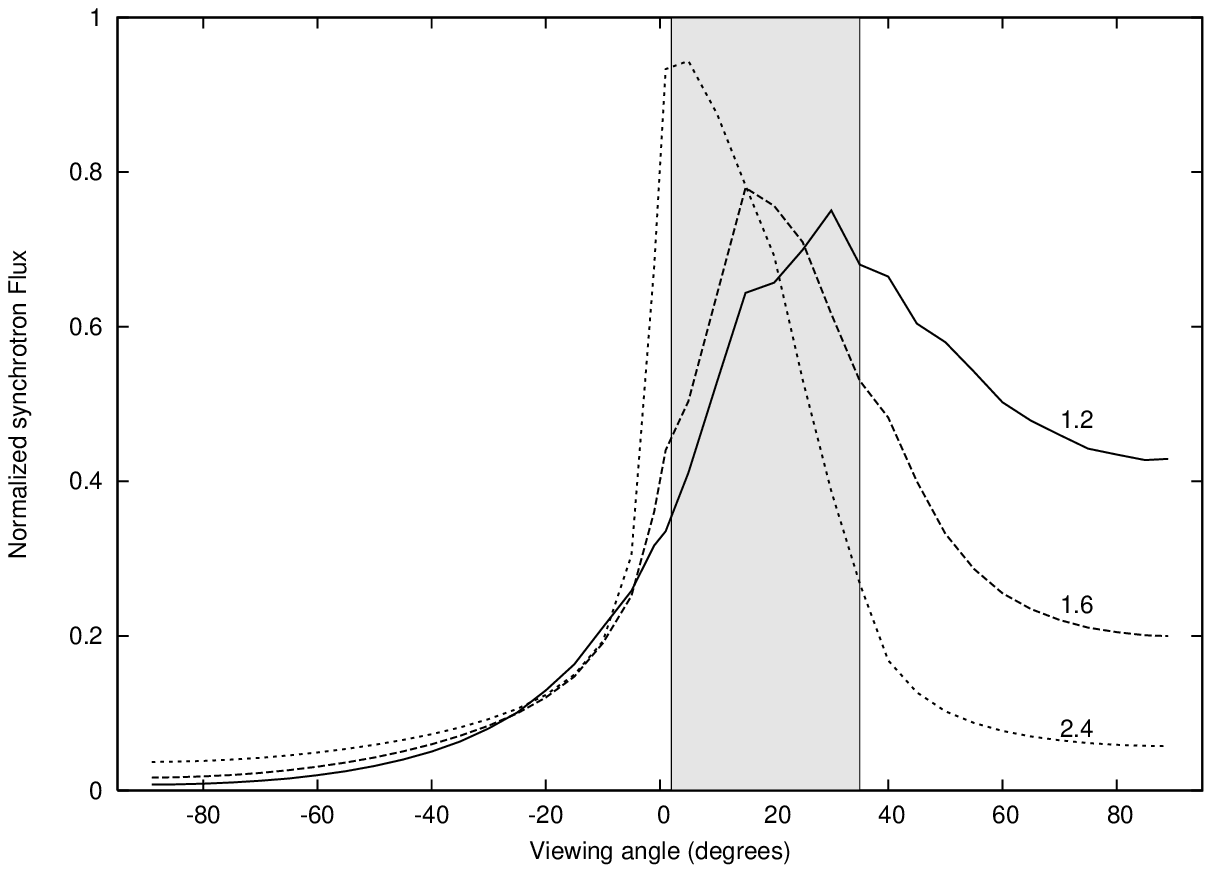,width=8.8cm}
\caption[]{a) The change in the synchrotron flux at 1.6~GHz as a function 
of viewing angle using the standard model parameters with 
$\zeta=10^{-4}$, but where $\Mdot_{\rm O}$ is adjusted to vary $\eta$. 
Shown are curves for $\eta=0.0316$ (solid), $\eta=0.1$ (dashed), and
$\eta=0.316$ (dotted). Each curve is normalized to the intrinsic 
synchrotron emission in each model. SSA, the Razin effect, and IC
cooling are included. The binary separation is $2 \times 10^{15} \cm$. 
At a viewing angle of
$-90^{\circ}$, the WR star is in front; at $0^{\circ}$ the lines of
sight are perpendicular to the axis of symmetry of the model; and at
$+90^{\circ}$ the O star is in front. Values for $R_{1}^{\rm O}/r_{\rm O}$
are noted next to each curve. The 
angular range of the WCR for the model with $\eta=0.316$ is shown
by the shaded region, with the asymptotic angles of the WR and O shocks 
delimiting its edges. b) As a) but where the binary separation (hence
$r_{\rm O}$) is
reduced by a factor of 4 to $5 \times 10^{14}\cm$.}
\label{fig:eta_vari}
\end{figure*}

In Fig.~\ref{fig:eta_vari}a we show how the synchrotron flux at 1.6~GHz
varies with viewing angle and $\eta$ when the binary separation is
$2 \times 10^{15} \;\cm$. The half-opening angle of the
WR shock, the CD, and the O shock as a function of $\eta$ are
noted in Table~\ref{tab:opening}\footnote{\cite{Eichler:1993} provide
an analytical expression for the half-opening angle of the CD,
$\theta_{\rm cd}$ (their Eq.~3). However, the statement that the angle
between the two shocks, $\Delta \theta \sim \theta$, is inaccurate, as
the data in Table~\ref{tab:opening} shows. Instead we find that
$\Delta \theta \approx 35-40^{\circ}$ for the range of $\eta$ noted in
Table~\ref{tab:opening}, whereas $\theta$ changes by almost a factor
of 2.}.  The main characteristic is an increase in absorption of the
synchrotron emission when the lines of sight to the apex of the WCR pass
close to either star.

\begin{table}
\begin{center}
\caption[]{Half-opening angles ($\theta$, in degrees, measured from the 
line of centres between the stars in our models), of the 
shocks in the WR and O 
winds bounding the WCR, and of the contact discontinuity, 
as a function of $\eta$. 
}
\label{tab:opening}
\begin{tabular}{llll}
\hline
\hline
$\eta$  & WR & CD & O \\
\hline
0.3160 & 88 & 69 & 55 \\ 
0.1000 & 74 & 52 & 36 \\
0.0316 & 56 & 36 & 17 \\
\hline
\end{tabular}
\end{center}
\end{table}

Here we vary $\eta$ by changing $\Mdot_{\rm O}$, so it is not
surprising to see that the absorption at $+90^{\circ}$ is highest when
$\eta$ is highest (i.e. $\Mdot_{\rm O}$ highest). When the O star is in
front, the amount of free-free absorption depends on the ratio of 
$R^{\rm O}_{1}/r_{\rm O}$ \citep{Eichler:1993}. At 1.6~GHz, this ratio 
is 0.58, 0.40, and 0.30 for $\eta=0.316$, 0.1, and 0.0316, respectively. Larger
values of $R^{\rm O}_{1}/r_{\rm O}$ result in higher absorption,
which is in agreement with Fig.~\ref{fig:eta_vari}a.  We naively
expected that the synchrotron emission would be least affected by
absorption when the observer looks down the WCR into the
system. However, we find that in all 3 cases, the absorption is a
minimum at viewing angles which are close to the asymptotic angle 
of the WR shock, and steadily increases as the viewing angle 
swings from being parallel to the WR shock to being parallel to the O shock.
We attribute this finding to the fact that even when the 
viewing angle is within the WCR, the far side of the
WCR is absorbed by the O star wind, and note that there are 
implications for any observing strategy attempting
to detect radio bursts in CWB systems (e.g., in $\gamma^{2}$~Velorum).

It is also interesting that the model with $\eta=0.316$ shows the
lowest relative absorption when the viewing angle is $-90^{\circ}$.  We
attribute this to the wider opening angle of the WCR in the model, and
the fact that more of the WCR is visible when $\eta$ is high.
Though not shown here, the variation with viewing angle is reduced at
higher frequencies, since $R_{1} \propto \nu^{-0.7}$.

The variation in synchrotron flux with viewing angle is more pronounced at
smaller binary separations, as shown in Fig.~\ref{fig:eta_vari}b,
where the values of $R^{\rm O}_{1}/r_{\rm O}$ are $4 \times$ larger
than previously. In all of the models in Fig.~\ref{fig:eta_vari}b,
the drop in emission as the sightlines move out of the WCR is steeper
on the WR side. 

If an observer's line of sight into the system lies in the orbital
plane, two bursts of radio emission will be seen as the WCR sweeps
past, as suggested by \cite{Eichler:1993}. We find that this occurs if
$R^{\rm O}_{1}/r_{\rm O} \gtsimm 0.4$. When $R^{\rm O}_{1}/r_{\rm
O} \gtsimm 2$, the burst is confined to a range in viewing angle equal
to $\Delta \theta$. However, if the inclination of the orbital
plane is such that the line of sight intercepts the limb of the WCR,
only one such burst per orbit may be seen. In contrast, the 
behaviour of the system with $\eta = 0.0316$ shown in 
Fig.~\ref{fig:eta_vari}a is better characterized as a single decline
in synchrotron emission when the WR star is in front of the O star,
since the emission remains high even when the O star is directly
in front of the WR star. 

The complexity demonstrated in Fig.~\ref{fig:eta_vari} further
illustrates the need for models which take into account the spatial
extent of the emission and absorption.


\begin{table*}[ht]
\begin{center}
\caption[]{Observations of \object{\wr147} from the literature.\label{tab:147fluxes}}
\begin{tabular}{llllll}
\hline
\hline
Freq.&Skinner$^{a,1}$&SetiaGunawan$^b$&Contreras$^c$&Churchwell$^d$&Average\\
(GHz)& (mJy)& (mJy)&(mJy)&(mJy)&(mJy)\\
\hline
$0.353$ &$-$         &$16.0\pm4.0$&$-$           &$-$           &$16.0\pm4.0$\\
$1.42$  &$25.2\pm0.4$&$26.4\pm2.0$&$-$           &$26.1\pm0.1^3$&$25.9\pm2.0$\\
$4.86$  &$37.3\pm1.6$&$35.4\pm0.4$&$38.4\pm0.1$  &$37.5\pm1.0^2$&$37.2\pm1.6$\\
$8.3$   &$42.5\pm2.2$&$-$         &$41.0\pm3.7^2$&$-$           &$41.8\pm3.7$\\
$14.94$ &$47.9\pm1.7$&$-$         &$57.4\pm0.3$  &$47.4\pm3.2^2$&$50.9\pm3.2$\\
$22.46$ &$54.7\pm2.7$&$-$         &$-$           &$59.1\pm1.2^3$&$56.9\pm2.8$\\
$43.0$  &$82.8\pm1.2$&$-$         &$82.8\pm1.1$  &$-$           &$82.8\pm4.1$\\
\hline
\end{tabular}
\end{center}

References: $a)$ \citet{Skinner:1999}, $b)$ \citet{SetiaGunawan:2001b}, $c)$
\citet{Contreras:1997}, and \citet{Contreras:1999} $d)$ \citet{Churchwell:1992}\\ 
Notes:\\ $^1$ \citet{Skinner:1999} quote two values for flux using
two different techniques on the same data. The value given here is the 
average value, along with the standard deviation.\\ 
$^2$ Error based on the standard deviation of multiple observations.\\
$^3$ Error assumed to be the uncertainty in the images.
\end{table*}

\section{The radio spectrum of \object{\wr147}}
\label{sec:wr147}

Having explored how the radio flux is affected by IC cooling, binary
separation, the wind momentum ratio, and variations in the B-field,
we once again turn our attention to the modelling of a specific system. 
\object{\wr147} is one of the brightest CWB systems, and is also one of the
systems where the stellar wind of the WR star and the WCR have been
resolved at radio wavelengths \citep{Williams:1997}. In addition,
\object{\wr147} has one of the most extensively observed radio spectra 
of any of the CWB systems, with observations from 353 MHz to 43 GHz.

In Paper I, we demonstrated a number of fits of our CWB models to the
synchrotron spectrum of \object{\wr147}. However, those models suffer
from two main problems. Firstly, the optically-thick turnover at
$\sim 1$~GHz was attributed to SSA which we
have discovered subsequently was miscalculated (see earlier footnote),
and a new fit with the corrected model code is now required. Secondly,
the synchrotron spectrum we were attempting to fit in Paper~I had been
derived from the observed total fluxes and an extrapolated thermal
spectrum for the stellar wind emission based on the flux of the WR star
stellar wind at several high frequencies
\citep[e.g.,][]{SetiaGunawan:2001b}.  Though a reasonable approach, the
resulting synchrotron spectrum exhibits considerable variation
dependent on the adopted thermal model. In an attempt to address this
problem, here we aim to fit the observed total radio emission at each
of the observed frequencies, and show the contributing synchrotron and
thermal spectra from the model.

Observations of \object{\wr147} are available from a number of
authors. These are summarised in Table~\ref{tab:147fluxes}.
\citet{Skinner:1999} used two methods to obtain flux estimates from the
same data, and these show considerable scatter. We have taken the liberty of 
quoting an average of the values from that paper. It appears that
\object{\wr147} is variable at some frequencies
\citep{SetiaGunawan:2001b}, so where observations at multiple epochs
are available \citep[e.g.,][]{Churchwell:1992, Contreras:1997,
Contreras:1999}, we have quoted an average flux. To attain a
reasonable estimate of the uncertainty in the flux values, we quote
either the variance of the observations (where there are multiple
observations), or an estimate of the uncertainty in the absolute flux
scale, whichever is largest. For the uncertainty in the flux scale, we
adopt 5\% for frequencies 15~GHz and greater, and 2\% for frequencies
less than 15~GHz \citep{Perley:2003}. We believe that our approach to
estimating the observed fluxes from \object{\wr147} is very
conservative, but it at least makes an attempt at assessing the
relative merits of all observations in the literature, and takes into
account the impact of variability.

\object{\wr147} is a very wide system with a projected separation
$D_{\rm sep}\cos~i=0.635\pm0.020$\arcsec \citep{Williams:1997}, where
$i$ is the angle at which we view the system.  At the estimated
distance of $\sim0.65$~kpc \citep{Churchwell:1992, Morris:2000} this
corresponds to a separation $D_{\rm sep}\sim415/\cos~i$~AU. As noted
in Paper~I, this relationship between $D_{\rm sep}$ and $i$ represents
an important constraint for any models of the system. We use the
stellar wind parameters from Paper I: $\Mdot_{\rm WN8} = 2 \times
10^{-5} \Msolpyr$, $v_{\rm \infty WN8} = 950 \kmps$, $v_{\rm \infty
OB} = 1000 \kmps$ and $\Mdot_{\rm OB} = 3.8 \times 10^{-7} \Msolpyr$,
giving a wind momentum ratio $\eta = 0.02$ \citep{Pittard:2002b}.
Of these parameters, $\Mdot_{\rm OB}$ is the most uncertain. This is
partly due to the considerable uncertainty in the companion's spectral
type - estimates range from B0.5 \citep{Williams:1997} to O5-7 I-II
\citep{Lepine:2001}. The luminosity of the companion ($L_{\rm OB}$)
ranges from $\sim 10^{5}-10^{6} \;\Lsol$, resulting in considerable
scope for varying the degree of IC cooling in the models.  The
composition of the WN8 stellar wind is taken from~\citet{Morris:2000},
giving $X=0.09$, $Y=0.89$, and $Z=0.016$. The wind temperature for
both stellar winds was assumed to be {10~kK, and the dominant
ionization states were H$^+$, He$^+$ and CNO$^{2+}$.

\begin{figure}
    \centering
    \includegraphics[angle=0,width=3.5in]{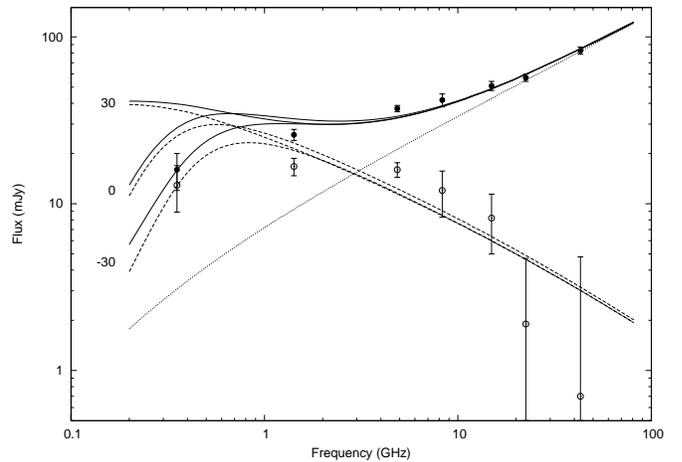}
    \caption{Model spectra for \object{\wr147}, including total
    (solid), synchrotron (dashed), and thermal (dotted) flux.  The
    observations (solid circles) are from
    Table~\ref{tab:147fluxes}. The open circles are synchroton fluxes
    estimated by subtracting the thermal model from the total observed
    fluxes. Viewing angles of $-30^{\circ}$ (model A), $0^{\circ}$
    (model B), and $30^{\circ}$ (model C) are shown. All models
    used $\eta=0.02$, $\gamma_{\rm max}=10^{5}$, $L_{\rm OB}=10^{5}\;\Lsol$,
    $\zeta=7.66\times 10^{-3}$, and a wind filling
    factor, $f$, of 0.139. The spectral index of the non-thermal electrons,
    $p$, was assumed to be 2. A slight curvature to the optically thin
    part of the synchrotron
    spectrum (\cf Fig.~\ref{fig:stan_spec1}) arises from IC 
    cooling.\label{fig:147_equipmod}}
\end{figure}

Firstly, we wish to repeat the modelling done in Paper I to evaluate
the impact of the miscalculated SSA. The resulting models are shown in
Fig.~\ref{fig:147_equipmod}. As mentioned earlier, SSA has little
impact at the frequencies of the observations in these systems, and we
have to turn to other mechanisms to generate the optically thick
turnover. With $\zeta_{\rm B}=\zeta_{\rm rel}$, we find that the low frequency
turnover is controlled by free-free absorption along the line-of-sight
through the circum-binary stellar wind envelope, which is contrary to
the conclusion from Paper I.  The line-of-sight opacity changes with
the viewing angle of the model: positive and negative viewing angles pass
first through the O star wind and the denser WR star wind
respectively. As a result, large negative angles are more opaque than
large positive values, resulting in lower observed fluxes at low
frequencies with negative viewing angles.  The quality of the fits
in Fig.~\ref{fig:147_equipmod} is largely determined by the model emission
at 353~MHz, with little variation between models at all
other observed frequencies. The $-30^\circ$ model matches the 353~MHz
data the best, though we have made no attempt to adjust parameters in the 
model to improve any of the fits. Rather, models A-C are intended to 
demonstrate the impact of changing the viewing angle on the behaviour of the
low frequency spectrum. $\zeta_{\rm B}$ and 
$\zeta_{\rm rel}$ are held constant in models A-C, and their ratio is set 
to 1. The intrinsic synchrotron luminosity from model B is slightly higher
than that in models A and C because the stars in model B are closer together
and this model has a viewing angle of $0^{\circ}$.
The fit to the 353~MHz data point is very sensitive to the viewing angle, 
and once this point is fit moderately well, the goodness-of-fit of 
the spectral models is largely determined by the flux computed at 1.4 and
5~GHz, given the smaller uncertainties we have derived for the observed
fluxes at these two frequencies.
We have calculated further models at various viewing angles, and find
that we cannot obtain satisfactory fits to both the 1.4 and 5~GHz data points 
with $\zeta_{\rm B} = \zeta_{\rm rel}$. 

In an attempt to improve the spectral fit, we have calculated
numerous models designed to explore the relevant parameter space.
Firstly, we varied the wind momentum ratio in the range $0.02-0.012$
\citep{Pittard:2002b}. Since the half-opening angle of the WCR is only
weakly dependent on $\eta$ in this range \citep[$\theta_{\rm cd} \propto
\eta^{1/3}$, see][] {Eichler:1993}, for a given viewing angle there is
very little difference to the {\em shape} of the spectrum. The normalization
of the free-free and synchrotron emission changes slightly, but this
can be compensated for by adjusting the degree of clumping and the
value of $\zeta$, respectively.  We also performed calculations with
an increased amount of clumping in the undisturbed winds, with the
mass-loss rates of both stars adjusted downwards in order to obtain
the desired amount of free-free emission. Changing
the volume filling factor of the clumps from $f=0.139$ to $f=0.03$ produces
an insignificant change in the flux at 353~MHz.

To simulate crudely a decrease in acceleration efficiency at the 
shocks away from the line-of-centers between the stars, we have
also calculated models where the synchrotron emission and absorption
is limited to a region within $\pi r_{\rm O}/2$ of the axis of symmetry
of the WCR. While this distance is somewhat arbitrary, we
note that this change has only a small effect on the synchrotron
spectrum once it is renormalized (by increasing $\zeta$).

The basic problem with all of the models noted so far is their total
emission spectrum is too flat between 1.5 and 5~GHz (see 
Fig~\ref{fig:147_equipmod}). The fit in this frequency range can be improved
by shifting the low-frequency turnover in the
model synchrotron spectrum to a higher frequency.  As demonstrated in
Fig.~\ref{fig:147_equipmod}, fits to these two data points
cannot be achieved effectively with free-free opacity alone. The turnover
can be moved to a higher frequency by decreasing the magnetic energy
density relative to the relativistic energy density, which increases
the impact of the Razin effect (recall $\nu_{\rm R}\propto1/{\rm
B}$). We induce this reduction in the magnetic energy density by
setting $\zeta_{\rm B}\neq\zeta_{\rm rel}$, as described in
Sec.~\ref{sec:zeta_var}.  The resulting synchrotron spectra are shown
in Fig.~\ref{fig:147_razin}, where it is clear that the turnover due
to the Razin effect moves to higher frequencies as $\zeta_{\rm
B}/\zeta_{\rm rel}$ decreases. The best fit to the 1.4 and 5~GHz data
is attained with $\zeta_{\rm B}/\zeta_{\rm rel}=10^{-4}$, which also
gives the best fit to all other data points except that at 353~MHz,
which is poor. No attempt has been made to improve the fits in models
D-H by adjusting any other parameters, though $\zeta_{\rm B}$ and
$\zeta_{\rm rel}$ were adjusted to renormalize the spectrum in order
to keep the comparison simple. When we reduce $\zeta_{\rm B}$ the
synchrotron luminosity drops, and we must increase $\zeta_{\rm rel}$
in order to maintain a specific synchrotron luminosity. This has the
consequence that the value of $\zeta_{\rm rel}$ is very close to unity
in model H.  

\begin{figure}
    \centering
    \includegraphics[angle=0,width=3.5in]{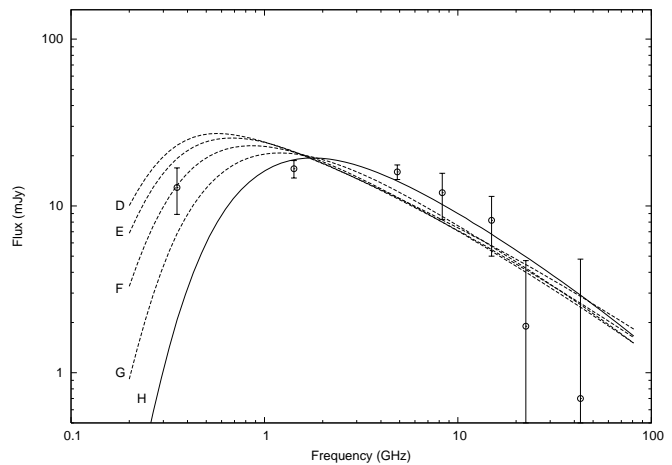}
    \caption{Model synchrotron spectra for \object{\wr147} at a
     viewing angle of $0^\circ$ and
     $\zeta_{\rm B}/\zeta_{\rm rel}=1,10^{-1}, 10^{-2}, 10^{-3}$ and $10^{-4}$
     (models D-H respectively, with $10^{-4}$ solid - see 
      Table~\ref{tab:models}). Data points
     are the same as in Fig~\ref{fig:147_equipmod}.\label{fig:147_razin}}
\end{figure}

\begin{table}
\label{tab:models}
\begin{center}
\caption[]{Summary of parameters used in our models of \object{\wr147} with
$p=2$ (models A-I) and $p=1.4$ (models J and K). A best fit to the data was 
only attempted with models I, J and K - models A-H are used to demonstrate 
the effect of varying the viewing angle or
$\zeta_{\rm B}/\zeta_{\rm rel}$. All models include
free-free absorption, SSA, the Razin effect, and IC cooling, and
were calculated with $\eta=0.02$ and $\gamma_{\rm max}=10^5$. Models A-I
had $f=0.139$ and $L_{\rm OB} = 10^{5}\;\Lsol$, model J had $f=0.170$ and
$L_{\rm OB} = 10^{5}\;\Lsol$, and model K had $f=0.157$ and
$L_{\rm OB} = 4 \times 10^{5}\;\Lsol$.}
\begin{tabular}{llll}
\hline
\hline
Model  &  Viewing angle ($^\circ$) &  ${\zeta}_{\rm B}/{\zeta}_{\rm rel}$ & $\zeta_{\rm rel}$ \\   
\hline
A      & -30  &  $1$         & $7.66\times10^{-3}$\\ 
B      &   0  &  $1$         & $7.66\times10^{-3}$\\ 
C      &  30  &  $1$         & $7.66\times10^{-3}$\\ 
D      &   0  &  $1$         & $7.27\times10^{-3}$\\ 
E      &   0  &  $10^{-1}$   & $2.00\times10^{-2}$\\ 
F      &   0  &  $10^{-2}$   & $5.69\times10^{-2}$\\ 
G      &   0  &  $10^{-3}$   & $1.72\times10^{-1}$\\ 
H      &   0  &  $10^{-4}$   & $5.57\times10^{-1}$\\ 
I      &  30  &  $10^{-4}$   & $5.56\times10^{-1}$\\ 
J      & -30  &  $1$         & $1.38\times10^{-2}$\\ 
K      & -60  &  $1$         & $1.54\times10^{-2}$\\ 
\hline
\end{tabular}
\end{center}
\end{table}

A better fit to the 353~MHz data point can be attained, in part, by
increasing the viewing angle to reduce the impact of free-free
opacity. We have attempted to fit the data by varying the viewing angle, 
$\zeta_{\rm B}$ and $\zeta_{\rm rel}$.
The resulting best fit is shown in Fig.~\ref{fig:best_fits}
(model I), and its weighted $\chi^{2}$ value is 10.7. With 4 degrees of
freedom, the reduced $\chi^{2}=2.7$. 
At the high frequency end the model can be made to more
closely fit the data points by reducing $\gamma_{\rm max}$, though
given the uncertainty in the data points, this has little impact on
the goodness of the model fit. 

\begin{figure}
    \centering
    \includegraphics[angle=0,width=3.5in]{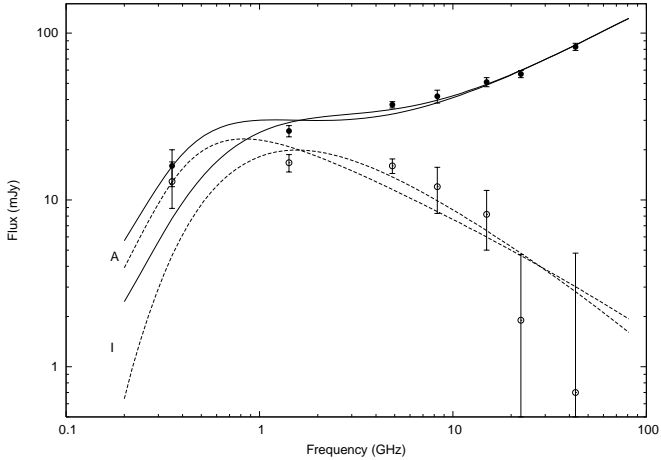}
    \caption{The total (solid) and synchrotron (dashed) spectra obtained
     with model I (see Table~\ref{tab:models}), where
     $\zeta_{\rm B}/\zeta_{\rm rel}=10^{-4}$. Also shown is model A for
     comparison. The thermal emission model is the same as used in 
     Fig.~\ref{fig:147_equipmod}.\label{fig:best_fits}}
\end{figure} 

A fundamental concern is the models suggest the DSA process in
\object{\wr147} is highly efficient, with an uncomfortably large
fraction of the pre-shock energy transferred to the non-thermal
electrons. While there are several ways in which the model assumptions
might lead us to overestimate $\zeta_{\rm rel}$ (e.g., the magnetic
field in the WCR may not be completely tangled resulting in
anisotropic emission; the thermal energy density of the WCR may have
been underestimated through our choice of mass-loss rates and/or wind
speeds, etc.), these possible solutions are either rather
unsatisfactory in themselves, and/or unlikely to substantially
increase $\zeta_{\rm B}/\zeta_{\rm rel}$. We note instead that in
situations where the particle acceleration is highly efficient, the
shock precursor (which influences the subshock compression, the
injection and acceleration efficiencies, and the shape of the
non-thermal particle energy spectrum) is modified by the back-pressure
of energetic particles \citep[e.g.,][]{Ellison:2004}. As noted in
Sec.~\ref{sec:nt_spectrum}, this can create a concave curvature to the
energy spectrum of the non-thermal particles, where $p$ decreases with
increasing energy. Two other effects can change the overall slope of
the energy spectrum. First, the softer equation of state of the
relativistic particles makes the shock plasma more compressible, and
increases the overall compression ratio, $r_{\rm tot}$ (when the
partial pressure from relativistic particles dominates the total
pressure of the plasma, $r_{\rm tot} \rightarrow 7$).  Since
$p=(r_{\rm tot}+2)/(r_{\rm tot}-1)$, a flatter power-law index results
($p=1.5$ in the limit that $r_{\rm tot}=7$). Second, the overall
compression ratio can become arbitrarily large as the highest energy
particles escape from the shocks \citep[see][for further
discussion]{Ellison:2004}, and in this limit $p \rightarrow 1$.

Since our modelling suggests that the non-thermal particle acceleration 
in \object{\wr147} is highly efficient, we have explored fits to the data 
where $p$ is reduced from its ``standard'' value of 2. With $\zeta_{\rm rel} =
\zeta_{\rm B}$ and the model parameters noted earlier in this
section, better fits to the total flux can be obtained with $p=1.4$.
The resulting fit (model J) is shown in Fig.~\ref{fig:wr147p}.
$\zeta=1.4 \times 10^{-2}$ indicates that the particle acceleration
is indeed efficient, justifying our adjustment of $p$.
A $p=1.4$ spectrum corresponds to an overall compression 
ratio of 8.5 ($r_{\rm tot}=(p+2)/(p-1)$), and is comparable to that
achieved in simulations of supernova remnants with highly efficient 
particle acceleration \citep{Ellison:2004}.
The high frequency synchrotron fluxes in the model can be reduced by 
increasing $L_{\rm OB}$, while maintaining a similar quality of fit 
(model K, see Fig.~\ref{fig:wr147p}), though we again emphasize that the 
fits to the \object{\wr147} data are to the {\em total} flux, not to model
deduced synchrotron fluxes which are very uncertain. 

An important point is that as the overall compression increases
($r_{\rm tot}>4$), the subshock compression ratio
declines. Since the subshock is responsible for heating the thermal
gas, the post-shock temperature drops below the value expected in the
strong-shock ideal-gas limit.  Analysis of the X-ray data from
\object{\wr147} indicates that the temperature of peak emission from
the thermal gas is $kT \approx 0.9 \keV$ \citep{Skinner:1999}. This
can be compared to the theoretical value, $kT = 1.96\mu v_{8}^{2} \keV$, 
expected after thermalization in the strong-shock
ideal-gas limit for material of mean particle mass $\mu m_{\rm H}$
flowing into a stationary shock at a typical wind terminal velocity
$\vinfty = 1000v_{8} \kmps$. As most of the X-ray emission should
arise from the shocked WN8 wind \citep[see][]{Pittard:2002c}, we set
$\mu = 1.17$ and $v_{8}=0.95$. The expected temperature is then $kT =
2.1 \keV$. While the contrast is not as large as that seen in
\object{\wr140}, where the theoretical temperature is 5 times higher
than what is observed \citep{Pollock:2005}, we nevertheless
conclude that there is perhaps some additional support for highly efficient
particle acceleration.

\begin{figure}
    \centering
    \includegraphics[angle=0,width=3.5in]{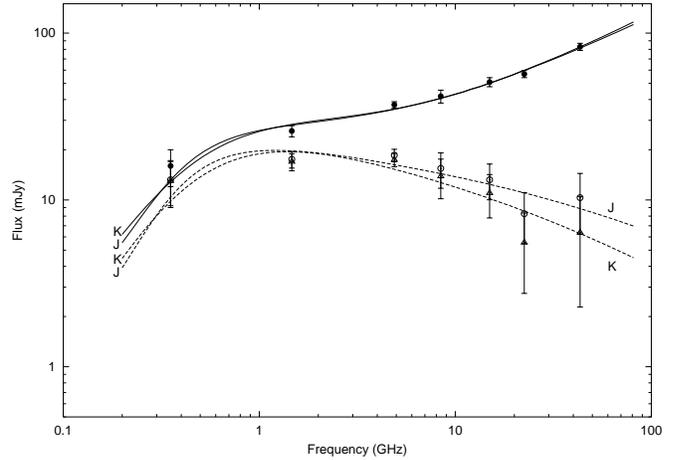}
    \caption{The total (solid), and synchrotron (dashed)
     spectra for \object{\wr147} with $p=1.4$ and $\eta=0.02$. Model~J has  
     $\zeta_{\rm B}=\zeta_{\rm rel}=1.4 \times 10^{-2}$, $f=0.17$, 
     $L_{\rm OB} = 10^{5}\;\Lsol$ and a viewing angle of $-30^\circ$.
     The weighted $\chi^{2}=4.97$, and the reduced $\chi^{2}=1.24$. Model~K
     has $\zeta_{\rm B}=\zeta_{\rm rel}=1.54 \times 10^{-2}$, $f=0.16$, 
     $L_{\rm OB} = 4 \times 10^{5}\;\Lsol$ and a viewing angle of $-60^\circ$.
     The weighted $\chi^{2}=5.25$, but there is only 2 degrees of freedom 
     and the reduced $\chi^{2}=2.63$.
     The open circles (triangles) are synchrotron fluxes estimated by
     subtracting the thermal fluxes in model~J (model~K) from the total 
     observed fluxes. The synchrotron data points are different between
     models J and K and from those in 
     Figs.~\ref{fig:147_equipmod}-\ref{fig:147_razin} because of the changes
     in the thermal flux by way of the filling factor.\label{fig:wr147p}}
\end{figure} 

\section{Summary and future directions}
\label{sec:summary}

We have extended the work in Paper~I by including the evolution of the
non-thermal electron energy distribution in the post-shock flow. IC cooling
is by far the most important process to consider, but we have also 
calculated the degree of ionic cooling for completeness.

We have demonstrated that the geometry of the WCR and the streamlines
of the flow within it lead to a spatially dependent break frequency in
the synchrotron emission. We therefore do not observe a single, sharp
break in the synchrotron spectrum integrated over the WCR, but rather
a steepening of the synchrotron spectrum towards higher
frequencies. 

As expected, IC cooling reduces the overall brightness of
the synchrotron emission from the WCR, but more so near the CD than
near the shocks. The synchrotron emission appears brighter at the WR
shock than at the O shock when the O star dominates the luminosity of
the system. In this situation, IC cooling has a somewhat lesser
effect at the WR shock.  Current radio interferometers do not have the
spatial resolution and sensitivity needed to detect such spatial
variation, but the predicted morphology may allow future observations
to determine whether the non-thermal electrons are primarily
accelerated at the shocks through the DSA scheme.

We find that the {\em observed} synchrotron emission may decrease with
decreasing $D_{\rm sep}$, through a combination of IC cooling, free-free
absorption, and the Razin effect.  This is consistent with the
non-detection of non-thermal emission from short period CWB systems.


We have also investigated how the synchrotron emission varies with the
assumed B-field, and have discovered that it is very difficult for SSA to
produce high turnover frequencies ($\gtsimm 1$~GHz). It is
much more likely that the cause of these turnovers is free-free
absorption through the circumbinary envelope, or the Razin effect.
If the optically thin {\em thermal} radio emission from the WCR, 
which scales as $D_{\rm sep}^{-1}$,
becomes comparable to the thermal emission from the unshocked winds,
it is possible to obtain a composite-like spectrum entirely from thermal
processes. 

Furthermore, the total free-free emission from CWB systems is likely
to vary with binary separation, though the exact nature of the
variation depends on the degree and distribution of clumping in the
system.  This has important consequences for deriving mass-loss rates
for stars in binary systems using radio fluxes, with the implication
that they may be significantly over- or under-estimated. Our models
show that it is preferable to use a high frequency ($> 15$~GHz) for
estimates of mass-loss rates.

We have applied our new model to \object{\wr147}, and find that the
introduction of IC cooling allows a better fit to the high frequency
data than obtained in Paper~I. However, since a weighted $\chi^2$
measure is biased by the relatively low uncertainity of the 1.4 and
5~GHz data points, any model which fits these points will be
preferred, and in this respect, the other data points are almost
irrelevant given their relative weights.
With the data currently available, we find that the best fits to the 1.4 
and 5~GHz fluxes are obtained when either the magnetic field energy
density is substantially less than the non-thermal electron energy density,
or when the injected non-thermal electron spectrum has a power-law index
of $p=1.4$. We find that we require efficient electron acceleration in 
both cases, which leads us to favour the model where $p=1.4$,
since non-linear calculations of DSA show a hardening of the non-thermal
particle spectrum. 

In summary, our analysis of the radio data from \object{\wr147}
indicates that efficient particle acceleration occurs in this
system, and that the shock structure may be modified by the back-pressure
of energetic particles. We emphasize that our current models are in
many ways very simplistic, and in order that these conclusions can be given a
firmer footing, more sophisticated models which self-consistently
include the back-reaction of the accelerated particles on the shock
structure need to be developed. Such calculations, unfortunately, are
not trivial, and a further complication is that the
shock obliquity may change with off-axis distance. As the obliquity
increases, the injection rates may reduce, so that some parts of the
shocks may be highly modified, while other parts will be unmodified.

Such self-consistent models would also be of more general use, because
despite the apparent efficiency of DSA, X-ray spectra from CWBs have
been modelled and interpreted assuming that the shocks place an
insignificant fraction of their energy into non-thermal particles
\citep[e.g.,][]{Stevens:1996,Zhekov:2000,Pittard:2002a}. These works have
led to inferences for important quantities such as the mass-loss
rates, wind speeds and composition, and the rate of
electron and proton equilibration.
However, the strong coupling between particle acceleration and thermal heating
implies that the inferences made from X-ray observations may differ
substantially between interpretations that include particle
acceleration self-consistently and those that do not. Future work
should address this issue and models should attempt to fit 
radio and X-ray data simultaneously. Finally, to further
improve our understanding of particle acceleration in \object{\wr147},
and other colliding wind binaries, accurate and precise radiometry 
is required in order to distinguish between models.





\begin{acknowledgements}
We would like to thank Don Ellison, Sam~Falle, Tom Hartquist, Sven Van
Loo, Perry Williams, and the anonymous referee for helpful
comments. JMP is supported by a University Research Fellowship from
the Royal Society.  This research has made use of NASA's Astrophysics
Data System Abstract Service.
\end{acknowledgements}

\appendix
\section{Synchrotron self-absorption (SSA)}
For an isotropic electron distribution function $f(p)$,
the absorption coefficient for SSA is given by 
\cite[see][Eq.~6.46]{Rybicki:1979}:
\begin{equation}\label{eq:SSA1}
\alpha_\nu = {{c^2}\over{8\pi h \nu^3}} \int 4\pi p^2 ~{\rm d} p~ \left(
f(p_1) - f(p) \right) P_\nu,
\end{equation}
where $P_\nu$ is the magneto-bremsstrahlung emissivity and
\begin{equation}
p_1^2 = (\gamma^2-1)(m_e c)^2 + (h\nu/c)^2 - 2h\nu \gamma m_e.
\end{equation}
Using $f(p) = {{n(E)}\over{4\pi p^2}} {{dE}\over{{\rm d}p}}$,
we find that
\begin{equation}\label{eq:SSA2}
\alpha_\nu = {{c^2}\over{8\pi h \nu^3}} \int \left(
{{n(E_1)}\over{p_1 E_1}} - {{n(E)}\over{p E}} \right) P_\nu p E~ {\rm d} E,
\end{equation}
where
\begin{equation}
E_1 = E - h\nu = \gamma m_e c^2 - h\nu.
\end{equation}
Re-writing the integral in Eq.~\ref{eq:SSA2} in terms of $\gamma$, we get
\begin{eqnarray}
\label{eq:SSA3}
& \alpha_\nu = & {{m_{e}^{2} c^5}\over{8\pi h \nu^3}} \times \nonumber \\
& &   \int \left(
{{n(\gamma-{{h\nu}\over{m_e c^2}})}\over{p_1 E_1}} - 
{{n(\gamma)}\over{p E}} \right) P_\nu 
\gamma \sqrt{\gamma^2-1} ~{\rm d} \gamma.
\end{eqnarray}
We evaluate Eq.~\ref{eq:SSA3} to find the SSA for any isotropic distribution
of particles $n(\gamma)$.



\label{lastpage}

\end{document}